\documentclass[12pt,twoside]{article}

\usepackage{a4wide}  
\usepackage{calc}
\usepackage{amsmath}
\usepackage{amsfonts,amssymb}
\usepackage[hyperindex]{hyperref} 
\usepackage{graphicx}

\usepackage{fancyhdr} 
\newcommand{\dL}{\frac{\partial}{\partial L}}
\newcommand{\ddL}{\frac{\partial^2}{\partial L^2}}
\newcommand{\dphi}{\frac{\partial}{\partial \varphi}}
\newcommand{\ddphi}{\frac{\partial^2}{\partial\varphi^2}}
\newcommand{\vph}{\varphi}
\newcommand{\deriv}[2]{\frac{\partial #1}{\partial #2}}

\newcommand{\ket}[1]{ \mid\! #1 \rangle}
\newcommand{\bra}[1]{ \langle #1 \!\mid}
\newcommand{\braket}[2]{ \langle #1 \mid #2 \rangle}
\newcommand{\scprod}[2]{ \langle #1 , #2 \rangle}
\newcommand{\expec}[1]{\left<#1\right>}
\newcommand{\U}{ \mathcal{\hat{U}}}
\newcommand{\D}{ \mathcal{D}}
\newcommand{\Tri}{ \mathcal{T}}
\newcommand{\Trans}{\mathbf{ \hat{T}}}
\newcommand{\Trace}{\mathrm{Tr}}
\newcommand{\Diff}{\mathrm{Diff}}
\newcommand{\Geom}{\mathrm{Geom}}
\newcommand{\Cat}{\mathrm{Cat}}
\newcommand{\Vol}{\mathrm{Vol}}
\newcommand{\genus}{{\mathfrak g}}
\newcommand{\mi}{\!-\!}
\newcommand{\equ}{\!=\!}
\newcommand{\pl}{\!+\!}
\newcommand{\ident}{\!\equiv\!}
\newcommand{\kl}{\!<\!}
\newcommand{\klgl}{\!\leqslant\!}
\newcommand{\gro}{\!>\!}
\newcommand{\grgl}{\!\geqslant\!}

\newcommand{\lin}{l_{in}}
\newcommand{\lout}{l_{out}}

\begin{document}

\title{\vspace{-1.5cm}
\Large \textbf{Analytic Results in 2D Causal Dynamical Triangulations: A Review\footnote{Based on the author's thesis for the Master of Science in Theoretical Physics, supervised by R.\ Loll and co-supervised by J.\ Ambj\o rn, J.\ Jers\'ak, July 2005.}} 
}
\author{{\large\em S. Zohren\footnote{Since 01.10.2005 at Blackett Laboratory, Imperial College, London, SW7 2AZ, UK.}}\\[10pt]
        {\footnotesize \rm Institute for Theoretical Physics, Utrecht University}\\[-5pt]
        {\footnotesize \rm Leuvenlaan 4, NL-3584 CE Utrecht, The Netherlands}\\
        {\footnotesize \rm and}\\
        {\footnotesize \rm Institut f\"ur Theoretische Physik E, RWTH-Aachen}\\[-5pt]
        {\footnotesize \rm D-52056 Aachen, Germany}\\[5pt]
        {\footnotesize \rm E-Mail: \href{mailto:zohren@physik.rwth-aachen.de}{zohren@physik.rwth-aachen.de}}\vspace*{-0.5cm}
 }
\date{}
\maketitle

\begin{abstract}

We describe the motivation behind the recent formulation of a nonperturbative path integral for Lorentzian quantum gravity defined through Causal Dynamical Triangulations (CDT). In the case of two dimensions the model is analytically solvable, leading to a genuine continuum theory of quantum gravity whose ground state describes a two-dimensional ``universe'' completely governed by quantum fluctuations. One observes that two-dimensional Lorentzian and Euclidean quantum gravity are distinct. In the second part of the review we address the question of how to incorporate a sum over space-time topologies in the gravitational path integral. It is shown that, provided suitable causality restrictions are imposed on the path integral histories, there exists a well-defined nonperturbative gravitational path integral including an explicit sum over topologies in the setting of CDT. A complete analytical solution of the quantum continuum dynamics is obtained uniquely by means of a double scaling limit. We show that in the continuum limit there is a finite density of infinitesimal wormholes. Remarkably, the presence of wormholes leads to a decrease in the effective cosmological constant, reminiscent of the suppression mechanism considered by Coleman and others in the context of a Euclidean path integral formulation of four-dimensional quantum gravity in the continuum. In the last part of the review universality and certain generalizations of the original model are discussed, providing additional evidence that CDT define a genuine continuum theory of two-dimensional Lorentzian quantum gravity. 
\end{abstract}

\newpage
\tableofcontents
\newpage
 
\section{Introduction}

Finding a consistent theory of quantum gravity which gives a fundamental quantum description of space-time geometry and whose classical limit is general relativity lies at the root of our complete understanding of nature. However, nearly one century has passed by since Einstein's invention of general relativity, and still very little is known about the ultimate structure of space-time at very small scales such as the Planck length. Because of the enormous energy fluctuation predicted by the Heisenberg uncertainty relation, the geometry at such short scales will have a non-trivial microstructure governed by quantum laws. The difficulty of finding a description of this microstructure is on the one hand due to the lack of experimental tests and further complicated by the fact that four-dimensional quantum gravity is perturbatively non-renormalizable. This also holds for supergravity or perturbative expansions in the string coupling in string-theoretic approaches.  One road to take is therefore to try to define quantum gravity \textit{nonperturbatively}.

Within those nonperturbative approaches of quantum gravity there are several attempts which suggest that the ultraviolet divergences can be resolved by the existence of a minimal length scale, commonly expressed in terms of the characteristic Planck length. A famous example is loop quantum gravity \cite{Thiemann:2002nj,Rovelli:2004tv,Nicolai:2005mc}; in this canonical quantization program the discrete spectra of area and volume operators are interpreted as an evidence for fundamental discreteness. Other approaches, such as four dimensional spin-foam models \cite{Perez:2003vx} or causal set theory \cite{Bombelli:1987aa,Sorkin:2003bx}, postulate fundamental discreteness from the outset. Unfortunately, neither of these quantization programs has succeeded so far in recovering the right classical limit.
 
A recent alternative is \textit{Lorentzian quantum gravity} defined through \textit{Causal Dynamical Triangulations} (CDT). In this path integral formulation a theory of quantum gravity is obtained as a continuum limit of a  superposition of space-time geometries. Thereby, in analogy to standard path integral formulations of quantum mechanics and quantum field theory, one uses an intermediate regularization through piecewise flat simplicial geometries.\footnote{However, whereas in quantum theory the piecewise straight paths are imbedded in a higher-dimensional space, this is \textit{not} the case for the piecewise flat geometries whose geometric properties are intrinsic.} 

Causal dynamical triangulations were first introduced in two dimensions \cite{Ambjorn:1998xu,Ambjorn:1999nc} as a nonperturbative path integral over space-time geometries with fixed topology which can be explicitly solved, leading to a continuum theory of two-dimensional Lorentzian quantum gravity. Later it was extended to three \cite{Ambjorn:2001cv,threed1,threed2} and four dimensions \cite{Ambjorn:2001cv,Ambjorn:2004pw,Ambjorn:2004qm,Ambjorn:2005db,Ambjorn:2005qt} (see also \cite{Ambjorn:2005jj} for a general overview). Among those recent developments there are very promising results regarding the large scale structure of space-time \cite{Ambjorn:2004pw,Ambjorn:2004qm}:  Firstly, the scaling behavior as a function of space-time volume is that of a genuine isotropic and homogeneous four-dimensional universe; this is the first step towards recovering a sensible classical limit. Moreover, after integrating out all dynamical variables apart from the scale factor $a(\tau)$ as a function of proper time $\tau$, it describes the simplest minisuperspace model used in quantum cosmology. 

Very interesting is the analysis of the microstructure of space-time: recent numerical results showed \cite{Ambjorn:2005db,Ambjorn:2005qt} that in this setting space-time does not exhibit fundamental discreteness. Instead, one observes a dynamical reduction of the dimension from four at large scales to two at small scales. This gives an indication that nonperturbative Lorentzian quantum gravity provides an effective ultraviolet cut-off through a \textit{dynamical dimensional reduction} of space-time. 

In this review we give a pedagogical introduction into CDT in two dimensions. The simple structure of two-dimensional gravity serves as a good playground to address fundamental concepts of the model which are also of significant relevance for the higher-dimensional realizations of CDT. 

One interesting fundamental question we want to address here is whether a sum over different space-time topologies should be included in the gravitational path integral. Since topology changes \cite{Dowker:2002hm} naturally violate causality, the sum over topologies is usually considered in Euclidean quantum gravity, where this issue does not arise. However, even in the simplest case of two-dimensional geometries the number of configurations contributing to the path integral grows faster than exponentially which makes the path integral badly divergent. Various attempts to solve this problem in the Euclidean context have so far been unsuccessful \cite{DiFrancesco:1993nw}, which leaves this approach at a very unsatisfactory stage.

Other attempts to define a path integral including a sum over topologies are either semiclassical or assume that a certain handpicked class of configurations dominates the path integral, without being able to check that they are saddle points of a full, nonperturbative description \cite{Hawking:2005kf}. However, we are not aware of any nonperturbative evidence supporting these ideas.\footnote{For another attempt to include the sum over topologies in the context of three-dimensional quantum gravity the reader is referred to \cite{Freidel:2002tg}.}

In the context of two-dimensional CDT it has recently been shown that one can unambiguously define a nonperturbative gravitational path integral over geometries and topologies \cite{Loll:2003rn,Loll:2005dr}\footnote{Also see \cite{Loll:2003yu,conf} for additional information.}. The idea of how to tame the divergences is to use certain causal constraints to restrict the class of contributing topologies to a physically motivated subclass. In the concrete setting these are geometrically distinguished space-times including an arbitrary number of infinitesimal ``wormholes'', which violate causality only relatively mildly. By this we mean that they do not necessarily exhibit macroscopic causality violations in the continuum limit. According to observable data this is a physically motivated restriction. This makes the nonperturbative path integral well defined. As an interesting result one observes that in the resulting continuum theory of quantum gravity the presence of wormholes leads to a decrease in the effective cosmological constant. This connects nicely to former attempts to devise a mechanism, the so-called \textit{Coleman's mechanism}, to explain the smallness of the cosmological constant 
in the Euclidean path integral formulation of {\it four}-dimensional quantum gravity 
in the continuum in the presence of infinitesimal wormholes \cite{Coleman:1988tj,Klebanov:1988eh}.

Another interesting aspect is the discussion of universality of two-dimensional Lorentzian quantum gravity defined through CDT. We provide explicit evidence for universality of the model which ensures that CDT defines a genuine continuum theory of two-dimensional quantum gravity and does not depend on the details of the regularization procedure.

The remainder of this review is structured a follows. In Section \ref{sec:CDT} we give an introduction to two-dimensional Lorentzian quantum gravity defined through CDT. We show that it gives rise to a well-defined continuum theory of two-dimensional quantum gravity, whose time evolution is unitary. Further, the restriction to the causal geometries in the path integral leads to a physically interesting ground state of a quantum ``universe'' which also shows that quantum gravity with Euclidean and Lorentzian signature are distinct theories. In Section \ref{sec:Topo} we address the question of how to incorporate the sum over topologies in the gravitational path integral. After restricting the contributing topologies to a physically motivated subclass including arbitrary numbers of infinitesimal wormholes, the path integral is well-behaved. The continuum theory can be obtained by means of a well-defined double scaling limit of the couplings. We observe a finite density of wormholes which leads to a decrease in the effective cosmological constant. In Section \ref{sec:univers} we show evidence for universality of the two-dimensional model. Further, we develop several one-to-one correspondences between two-dimensional CDTs and certain one-dimensional combinatorial structures, such as heaps of dimers and random walks. Appendices \ref{App:Lorentzian}-\ref{App:Calogero} provide supplementary information for Section \ref{sec:CDT}. In Appendix \ref{App:Lorentzian} we give a brief summary of the results on Lorentzian angles which naturally appear in the framework of CDT. In Appendix \ref{App:boundary} triangulations with different boundary conditions are discussed. In Appendix \ref{App:hamilton} we give an alternative derivation of the continuum quantum Hamiltonian obtained in Section \ref{sec:CDT}. Regarding the calculation of continuum quantities further details are given in Appendix \ref{App:Calogero}. In Appendix \ref{app:scaling} we discuss other double scaling limits related to the model in Section \ref{sec:Topo}, which we discarded as unphysical.

\section{2D Lorentzian quantum gravity}\label{sec:CDT}

In this section we give an introduction to two-dimensional Lorentzian quantum gravity defined through causal dynamical triangulations (CDT). The section is structured as follows: In Section \ref{sec:pathint} we introduce the reader to the concepts and problems of how to define a nonperturbative path integral for quantum gravity. In Section \ref{sec:geometry} an explicit definition of the gravitational path integral as a regularized sum over space-time triangulations is presented. The problem of how to incorporate Lorentzian space-times in this framework and how to give a well-defined prescription for a Wick rotation is discussed in Section \ref{sec:lorentziannature}; this leads to the notion of causal dynamical triangulations. In Section \ref{sec:transfer} a complete analytic solution to the discrete problem is presented. Performing the continuum limit of the discretized model, as done in Section \ref{sec:cont}, leads to a continuum theory of two-dimensional Lorentzian quantum gravity, which we will interpret in terms of its physical observables in Section \ref{sec:obs}.

\subsection{A  nonperturbative path integral for quantum gravity}\label{sec:pathint}

In this section a notion of a nonperturbative path integral for quantum gravity is introduced. In contrast to usual path integral formulations in quantum mechanics and quantum field theory the gravitational path integral is more involved due to the \textit{diffeomorphism-invariance} of the theory, which in terms of local coordinate charts are smooth invertible coordinate transformations $x^\mu\mapsto y^\mu(x^\mu)$, and due to the absence of a preferred metric background structure. 

The aim of this section is to expose this problem and to give first indications on how to define a gravitational path integral. Most of the arguments presented in this section can be found in \cite{Loll:2002xb}. For a more comprehensive account of path integrals in quantum field theory and the connection to critical phenomena the reader is referred to \cite{Zinn-Justin:2002ru}.

Before going into the conceptual details of the problems one encounters when defining a gravitational path integral, let us first recall the path integral representation of an one-dimensional (non-relativistic) quantum mechanical problem described by the time dependent Schr\"odinger equation
\begin{equation}
-i\frac{\partial}{\partial t}\ket{\psi,t}=\hat{H}(t)\ket{\psi,t}.
\end{equation}
The time evolution of the wave function is given by
\begin{equation}
\label{eq:path:timeevol}
\ket{\psi,t''}=\U(t'',t')\ket{\psi,t'},\quad \U(t,t)=\mathbf{1},
\end{equation}
where $\U(t,t')$ is defined by $\hat{H}(t)=-i\left[\partial_t \,\U(t,t') \right]_{t=t'}$. In the position representation, the time evolution, \eqref{eq:path:timeevol}, can be written as
\begin{equation}
\label{eq:path:timeevolpos}
\psi(x'',t'')=\int dx' \,G(x'',x';t'',t') \,\psi(x',t'),
\end{equation}
where $G(x'',x';t'',t')\equiv \bra{x''}\U(t'',t')\ket{x'}$ is the so-called propagator or Feynman kernel. For the case that the Hamiltonian $\hat{H}(t)$ is bounded from below, $\U(t,t')$ satisfies the semi-group property
\begin{equation}
\U(t'',t')=\U(t'',t)\,\U(t,t'),\quad t'\leqslant t\leqslant t''.\label{eq:path:semi}
\end{equation}

\begin{figure}[t]
\begin{center}
\includegraphics[width=3in]{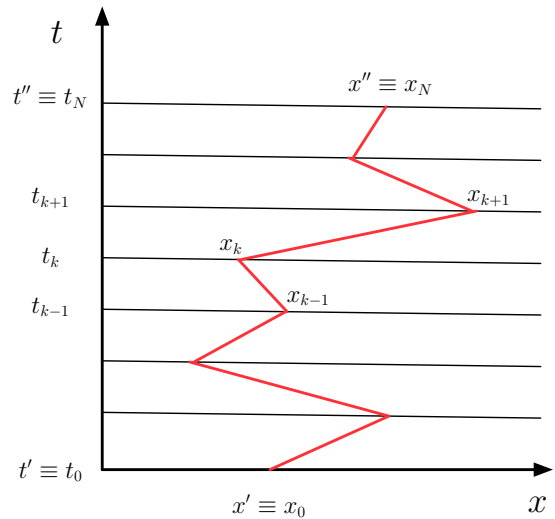}
\caption{Illustration of the path integral for a one-dimensional non-relativistic quantum mechanical problem, e.g. a propagating particle. One possible path of the configuration space (path space) is drawn. The ``virtual'' particle is propagating from $x_0$ to $x_N$ in a piecewise linear path of N steps of time $\epsilon=(t''-t')/N$ each.}
\label{fig:gravpathqm}
\end{center}
\end{figure}

This property \eqref{eq:path:semi} allows us to define the propagator as a limiting procedure of products of evolution operators corresponding to infinitesimal time intervals $\epsilon=(t''-t')/N$ (Figure \ref{fig:gravpathqm}),
\begin{eqnarray}
G(x'',x';t'',t') & = & \lim_{\epsilon\rightarrow 0} A^{-N}\prod_{k=1}^{N-1}\int dx_k\prod_{j=0}^{N-1}\bra{x_{j+1}}\U(t_{j+1},t_j)\ket{x_j} \nonumber\\
 & = &   \lim_{\epsilon\rightarrow 0} A^{-N}\prod_{k=1}^{N-1}\int dx_k \exp\left\{ i\sum_{j=0}^{N-1} \epsilon \mathcal{L}(x_{j+1},(x_{j+1}-x_{j})/\epsilon)\right\},\label{eq:path:pathint}
\end{eqnarray}
where $A$ is a normalization factor and $\mathcal{L}$ is the Lagrangian corresponding to $H(t)$. The integral in \eqref{eq:path:pathint} is taken over piecewise linear paths $x(t)$ from $x'\equiv x_0$ to $x''\equiv x_N$. The right hand side of \eqref{eq:path:pathint} is called the \textit{path integral} and is often written symbolically as
\begin{equation}
\label{eq:path:pathintsymb}
G(x'',x';t'',t')=\int_{\mathcal{P}} \D[x(t)] e^{i S[x(t)]},
\end{equation}
where $\D[x(t)]$ is a functional measure on the \textit{path space} $\mathcal{P}$ \cite{Roepstorff:1994ga}. The weight of each path is given by the classical action $S[x(t)]$.

Note that \eqref{eq:path:pathintsymb} is mathematically not well-defined. To actually perform the integral in \eqref{eq:path:pathintsymb} one usually does an analytic continuation on the time variable $t$ from the real to the imaginary axis, i.e. $t\mapsto \tau=it$. This so-called Wick-rotation makes the integrals real \cite{Reed:1975uy} by mapping $i\,S[x(t)]\mapsto -S^{eu}[x(t)]$, where $S^{eu}[x(t)]$ is the Euclidean action. After having performed the integration in the Euclidean sector one applies the inverse Wick rotation $\tau\mapsto-it$ to get back to the \textit{physically} meaningful results.

The question which arises now is: Can one define a path integral for quantum gravity using a similar strategy like the one used to arrive at \eqref{eq:path:pathintsymb}? The first step to answer this question is to find the analogue of path and the path space in a gravitational sense. One might think that the equivalent of the path space $\mathcal{P}$ in the case of the one-dimensional quantum mechanical problem is the space of all geometries in the gravitational case. To clarify this, let us become more precise and provide a definition of these concepts. Classically, a geometry is a space-time $(M,g)$, a smooth manifold $M$ equipped with a metric tensor field $g_{\mu\nu}$ with Lorentzian signature \cite{Wald:1984rg}. Due to the diffeomorphism invariance of the theory any two metrics are equivalent if they can be mapped onto each other by a coordinate transformation. Therefore all physical degrees of freedom are encoded in the equivalence class $[g_{\mu\nu}]=g_{\mu\nu}/\Diff(M)$ which we simply refer to as a \textit{geometry}. From this follows a natural notion of the \textit{space of all geometries} as the coset space $\Geom(M)=\mathrm{Metrics}(M)/\Diff(M)$.

\begin{figure}[h]
\begin{center}
\includegraphics[width=2.7in]{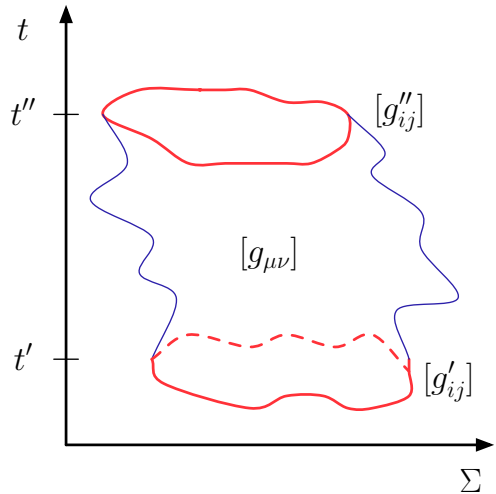}
\caption{Illustration of the gravitational path integral; one geometry of the configuration space is drawn. The geometry $[g_{\mu\nu}]$ has a time-sliced structure with respect to proper time $t$; it has an initial spatial geometry $[g_{ij}']$ at proper time $t'$ and final spatial geometry $[g_{ij}'']$ at proper time $t''$. Note that in the case of two dimensions the spatial hypersurface $\Sigma$ is one-dimensional and the spatial geometry $[g_{ij}]$ is totally characterized by its length $L$.}
\label{fig:gravpath}
\end{center}
\end{figure}

Having defined the space of all geometries we can use a foliation of the space of geometries to schematically define the gravitational path integral from an initial spatial geometry $[g_{ij}']$ at proper time $t'$ to a final spatial geometry $[g_{ij}'']$ at proper time $t''$ (Figure \ref{fig:gravpath}) by
\begin{equation}
\label{eq:path:gravpath}
G([g_{ij}''],[g_{ij}'];t'',t')=\int_{\Geom(M)} \D[g_{\mu\nu}] e^{i S_{\mathrm{EH}}[g_{\mu\nu}]}.
\end{equation}
Here, $\D[g_{\mu\nu}]$ is the functional measure on the space of all geometries $\Geom(M)$ and each geometry is weighted by the classical Einstein-Hilbert action $S_{\mathrm{EH}}[g_{\mu\nu}]$.\footnote{We assume that each geometry is weighted by the classical Einstein-Hilbert action including a term with a positive cosmological constant. One might also include higher curvature terms in the action. However, as we will see in Section \ref{sec:higher}, the resulting continuum theory belongs to the same universality class as the continuum theory obtained from $S_{\mathrm{EH}}$.} Before trying to give a concrete meaning to \eqref{eq:path:gravpath} in the next section, let us first discuss the conceptional problems of how to define and actually evaluate \eqref{eq:path:gravpath}. 

First of all, $\D[g_{\mu\nu}]$ has to be defined in a covariant way to preserve the diffeomorphism invariance. Since there is no obvious way to parametrize geometries, one would have to introduce covariant metric field tensor. Unfortunately, to perform further calculations one would then have to gauge fix the field tensors which would give rise to Faddeev-Popov determinants \cite{Faddeev:1980be} whose nonperturbative evaluation is exceedingly difficult.\footnote{See \cite{Distler:1988jt,Mottola:1995sj} for an evaluation in the setting of two-dimensional Euclidean quantum gravity in the light-cone gauge. In \cite{Dasgupta:2001ue} a calculation for three- and four-dimensional Lorentzian quantum gravity in the proper-time gauge is presented. It is anticipated there that the Faddeev-Popov determinants cancel the divergences coming from the conformal modes of the metric nonperturbatively.}

 The second problem is due to the complex nature of the integrand. In quantum field theory this problem is solved by doing the Wick rotation $t\mapsto \tau=it$, where $t$ is the time coordinate in Minkowski space $(t,\vec{x})$. Clearly, a prescription like this does not work out in the gravitational setting, since all components of the metric field tensor depend on time. Further $t\mapsto \tau=it$ is certainly \textit{not} diffeomorphism invariant (as a simple example consider the coordinate transformation $t\mapsto t^2$ for $t>0$). Hence, the question arises: What is the natural generalization of the Wick rotation in the gravitational setting? 
 
Finally, since we are working in a field theoretical context, some kind of regularization and renormalization will be necessary, and again, has to be formulated in a covariant way.
In quantum field theory the use of \textit{lattice} methods provides a powerful tool to perform nonperturbative calculations, where the lattice spacing $a$ serves as a cutoff of the theory. An important question to ask at this point is whether or not the theory becomes independent of the cutoff. Consider for example QCD and QED on the lattice. All evidence suggests that QCD in four space-time dimensions is a genuinely continuum quantum field theory. By genuine continuum quantum field theory we mean a theory which needs a cutoff at an intermediate step, but whose continuum observables will be \textit{independent} of the cutoff at arbitrary small scales. In QED the situation seems to be different, one is not able to define a non-trivial theory with the cut-off removed, unless the renormalized coupling $e_{ren}=0$ (trivial QED). Therefore, QED is considered as a low-energy effective theory of an elaborate theory at the Planck scale. However, as we will in Section \ref{sec:cont}, for the case of the gravitational path integral the continuum limit exists and the resulting theory is independent of the cutoff, in the same way as QCD is.

\subsection{Geometry from simplices}\label{sec:geometry}

As motivated in the previous section, regularization through lattice methods might be a powerful tool for a nonperturbative path integral formulation of quantum gravity. 
This obviously requires some kind of discretization of the geometries. 

The use of discrete approaches to quantum gravity has a long history \cite{Loll:1998aj}. In classical general relativity, the idea of approximating a space-time manifold by a \textit{triangulation} of space-time goes back to the early work of Regge \cite{Regge:1961px} and was first used in \cite{Rocek:1982fr,Frohlich:1994gc} to give a path integral formulation of gravity. By a triangulation we mean a piecewise linear space-time obtained by a gluing of simplicial building blocks. This one might think of as the natural analogue of the piecewise linear path we used to describe the path integral of the one-dimensional quantum mechanical problem. In two dimensions these simplicial building blocks are \textit{flat} Euclidean or Minkowskian triangles, where flat means isomorphic to a piece of Euclidean or Minkowskian space respectively. One could in principle assign a coordinate system to each triangle to recover the metric space $(M,g_{\mu\nu})$, but the strength of this ansatz lies just in the fact that even without the use of coordinates, each geometry is completely described by the set of edges length squared $\{l_i^2\}$ of the simplicial building blocks. This provides us with a regularized parametrization of the space of all geometries $\Geom(M)$ in a diffeomorphism invariant way, and hence, is the first step towards defining the gravitational path integral \eqref{eq:path:gravpath}.

\begin{figure}[t]
\begin{center}
\includegraphics[width=4.5in]{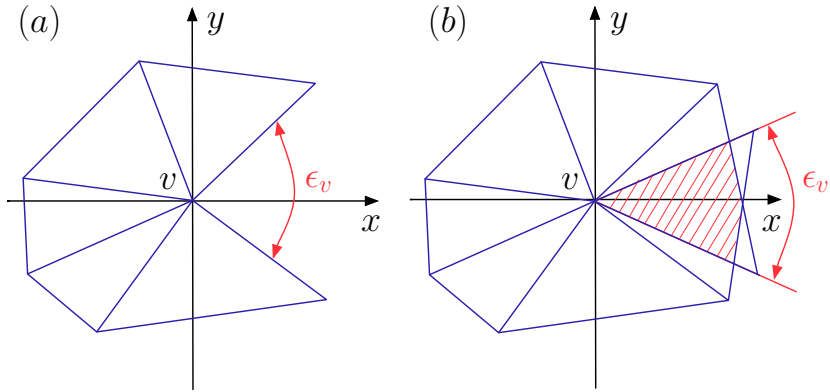}
\caption{Illustration of a positive (a) and negative (b) space-like deficit angle $\epsilon_v$ at a vertex $v$.}
\label{fig:curvature}
\end{center}
\end{figure}

For the further discussion it is essential to understand how a geometry is encoded in the set of edges length squared $\{l_i^2\}$ of the simplicial building blocks of the corresponding triangulation. In the case of two dimensions this can be easily visualized, since the triangulation consists just of triangles. Further, the Riemann scalar curvature $R(x)$ coincides with the Gaussian curvature $K(x)$ up to a factor of $1/2$. There are several ways to reveal curvature of a simplicial geometry. The most convenient method is by parallel transporting a vector around a closed loop. Since all building blocks are flat, a vector parallel transported around a vertex $v$ always comes back to its original orientation unless the angles $\theta_i$ of the surrounding triangles do not add up to $2 \pi$, but differ by the so-called \textit{deficit angle} $\epsilon_v=2\pi-\sum_{i\supset v}\theta_i$ (Figure \ref{fig:curvature}). The Gaussian curvature located at the vertex $v$ is then given by
\begin{equation}
\label{eq:simpl:gauss}
K_v=\frac{\epsilon_v}{V_v},
\end{equation}
where $V_v$ is the volume associated to the vertex $v$; more precisely, the dual volume of the vertex. Note that the curvature at each vertex takes the form of a conical singularity.
Then one can write the simplicial discretization of the usual curvature and volume terms appearing in the two-dimensional Einstein-Hilbert action,
\begin{eqnarray}
\frac{1}{2}\int_M d^2x \sqrt{|\det g|}\, R(x) & \rightarrow & \sum_{v\in \mathcal{R}} \epsilon_v, \label{eq:simpl:eps}\\
\int_M d^2x \sqrt{|\det g|}  & \rightarrow &  \sum_{v\in \mathcal{R}} V_v ,\label{eq:simpl:A}
\end{eqnarray}
where $\mathcal{R}\equiv \{l_i^2\}$ is a triangulation of the manifold $M$ described by the set of edge length squared. From this one can write down the simplicial discretization of the two-dimensional Einstein-Hilbert action, the so-called Regge action,
\begin{equation}\label{eq:simpl:reggeaction}
S_{\mathrm{Regge}}(\mathcal{R})=  \sum_{v\in \mathcal{R}} V_v \left(\lambda-k\,\frac{\epsilon_v}{V_v} \right),
\end{equation}
where $k$ is the inverse Newton's constant and $\lambda$ the cosmological constant. It is then an easy exercise of trigonometry to evaluate the right hand side of \eqref{eq:simpl:reggeaction} in terms of the squared edge lengths.

The treatment explained above can also be generalized to an arbitrary dimension $d$ in a straightforward way \cite{Miller:1997wb}. 

From this notion of triangulations one can now write the gravitational path integral as the integral over all possible edge lengths $\int\mathcal{D}l$, where each configuration $\mathcal{R}$ is weighted by the corresponding Regge action \eqref{eq:simpl:reggeaction}. A potential problem one encounters with this ansatz is an overcounting of possible triangulations, due to the fact that one can continuously vary each edge length. Further, one still has to introduce a suitable cut-off for the length variable $l$. Among other things, this motivated to the approach of ``rigid'' Regge calculus or \textit{dynamical triangulations} where one considers a certain class $\Tri$ of simplicial space-times as an explicit, regularized version of $\Geom(M)$, 
where each triangulation $T\in\Tri$ only consists of simplicial building blocks whose space-like edges have all the same edge length squared $l_s^2=a^2$ and whose time-like edges have all the same edge length squared $l_t^2=-a^2$. Here, the geodesic distance $a$ serves as the short-distance cutoff, which will be sent to zero later. At this point it is important to notice that fixing the edge length squared is not a restriction on the metric degrees of freedom. One can still achieve all kinds of deficit angles of either sign or directions by a suitable gluing of the simplicial building blocks.

From this one is able to give a definite meaning to the formal continuum path integral \eqref{eq:path:gravpath} as a discrete sum over inequivalent triangulations,
\begin{equation}
\label{eq:simpl:pathsum}
\int_{Geom(M)} \D[g_{\mu\nu}] e^{i S_{\mathrm{EH}}[g_{\mu\nu}]} \rightarrow \sum_{T\in\Tri} \frac{1}{C(T)} e^{i S_{\mathrm{Regge}}(T)},
\end{equation}
where $1/C(T)$ is the measure on the space of discrete geometries, with $C(T)=|\mathrm{Aut}(T)|$ the dimension of the automorphism group of the triangulation $T$. 

\subsection{Lorentzian nature of the path integral}\label{sec:lorentziannature}

After we have given an explicit meaning to gravitational path integral as a sum over inequivalent triangulations \eqref{eq:simpl:pathsum}, there still remains the problem of how to actually perform the sum, since it is up to now unclear what the prescription of a Wick rotation looks like, i.e. how we are going to analytically continue to the Euclidean sector of the theory.

To avoid this problem one first considered dynamical triangulations \cite{Ambjorn:1997di,Ambjorn:1999nc} on Euclidean geometries made up of Euclidean simplicial building blocks, where one does the \textit{ad hoc} substitution
\begin{equation}
\label{eq:lorentz:adhoc}
\int_{\Geom(M^{lor})} \D[g^{lor}_{\mu\nu}] e^{i S[g^{lor}_{\mu\nu}]}\rightarrow
\int_{\Geom(M^{eu})} \D[g^{eu}_{\mu\nu}] e^{- S[g^{eu}_{\mu\nu}]},
\end{equation}
where one simply replaced by hand the Lorentzian geometries with their Euclidean counterparts.  
The reason for doing so was not necessarily that one thought that Euclidean manifolds are more fundamental than Lorentzian manifolds, it was just that one had no \textit{a priori} prescription for a Wick rotation.

The potential problem with the substitution \eqref{eq:lorentz:adhoc} is that one integrates over geometries which know nothing about time, light cones and causality. Hence, there is no a priori prescription of how to recover causality in the full quantum theory. It is unlikely that this can be done by just performing the inverse Wick rotation $t\mapsto -it$. Another thing which fails is that $\Geom(M^{eu})$ has too many configurations and one integrates over a large class of highly degenerate \textit{acausal} geometries. This problem exists in higher dimensions $d\geqslant 3$, where it causes the absence of a well defined continuum theory. More precisely, due to the degenerate geometries, Euclidean dynamical triangulations in $d=3,4$ has an ``unphysical'' phase structure, where neither of the phases seems to have a ground state that resembles an extended geometry.\footnote{See \cite{Loll:1998aj} and references therein for a detailed discussion of the phase structure of four-dimensional dynamical triangulations.} 

This led to the approach of \textit{Lorentzian dynamical triangulations} or \textit{causal dynamical triangulations} (CDT) which was first introduced in two dimensions in \cite{Ambjorn:1998xu}, further elaborated in \cite{Ambjorn:1999nc,Ambjorn:2001cv} and has recently produced interesting results in $d=4$ \cite{Ambjorn:2005qt,Ambjorn:2005db,Ambjorn:2004pw,Ambjorn:2004qm}, as presented in the introduction. The main point of CDT is to insist in taking the Lorentzian structure, i.e. the inherent light-cone structure, seriously from the outset. Due to the triangulated structure one is able to find a well-defined Wick rotation on the full path integral \eqref{eq:simpl:pathsum} as we will discuss in the following. Once having defined the Wick rotation all calculations, including the continuum limit, are performed in the Euclidean sector, where at the end one goes back to the physical signature by performing the inverse Wick rotation (in the continuum theory).

Due to the inherent causal structure, the Lorentzian model is generally inequivalent to the Euclidean model.\footnote{In two dimensions the Euclidean model can be related to the Lorentzian one by a non-trivial procedure of integrating out baby universes \cite{Ambjorn:2001cv}.\label{foot:EuLor}} One might think of the causal structure as being equivalent to choosing a measure on the path integral \eqref{eq:simpl:pathsum} which suppresses acausal geometries and therefore also some of the highly degenerate geometries.

To be able to define the Wick rotation in this context, let us first clarify which are the \textit{causal} triangulations $T$ that contribute to the sum in the path integral \eqref{eq:simpl:pathsum}. 

\begin{figure}[t]
\begin{center}
\includegraphics[width=3in]{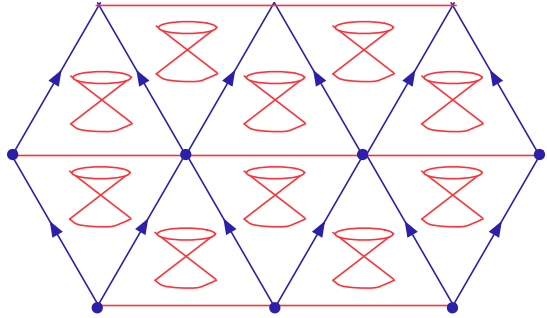}
\caption{Illustration of the causal structure in a flat triangulated manifold. The triangles are glued according to the sliced structure. Due to the inherent causal structure of the Minkowskian triangles, the whole simplicial manifold has a causal structure.}
\label{fig:pastfuture}
\end{center}
\end{figure}

We consider ``globally hyperbolic'' simplicial manifolds with a sliced structure, where the one-dimensional spatial hypersurfaces
have fixed topology of $S^1$; this will lead to a de-Sitter space-time in the continuum theory.  Further we will not allow for topology changes of the spatial slices; in how far topology changes can be incorporated within this framework will be discussed in Section \ref{sec:Topo}. The sliced structure of the simplicial manifold provides us with a preferred notion of ``proper time'', namely, the parameter labeling successive spatial slices. Note that this use of time is not a gauge choice, since proper time is naturally defined in a diffeomorphism invariant way.  
Several spatial slices are then connected by Minkowskian triangles\footnote{A brief description of how to assign angles and volumes to Minkowskian triangles is given in Appendix \ref{App:Lorentzian}.} with one space-like edge of length squared $l_s^2=+a^2$ and two time-like edges of length squared $l_t^2=-a^2$. Since all simplicial building blocks are Minkowskian triangles which have an intrinsic light-cone structure, we have local causality relations which, if one uses certain gluing rules, imply a global causal structure on the whole manifold (Figure \ref{fig:pastfuture}).

From this notion, one is now able to define the Wick rotation on a very elementary geometric level. By virtue of the triangulated structure one can define a Wick rotation on each simplicial building block respectively by performing a Wick rotation on all time-like edges $l_t$ of the Lorentzian triangles, i.e. $l_t\mapsto i \,l_t$. The Wick rotation acting on the whole simplicial manifold $\mathbb{T}$ is then defined by the following injective map
\begin{equation}
\label{eq:simpl:Wick}
\mathcal{W}:\quad\mathbb{T}^{lor}=\{T,{l_s^2=a^2,l_t^2=-a^2}\}\mapsto
\mathbb{T}^{eu}=\{T,{l_s^2=a^2,l_t^2=a^2}\}.
\end{equation}
In Appendix \ref{App:Lorentzian} it is shown how to implement $\mathcal{W}$ in a mathematically clean way. Further, it is shown there that in the path integral, the Wick rotation $\mathcal{W}$ implements precisely the desired analytic continuation on the Regge action,
\begin{equation}
\mathcal{W}:\quad  e^{i\,S_{\mathrm{Regge}}(T^{lor})}\mapsto  e^{-\,S_{\mathrm{Regge}}(T^{eu})}.
\end{equation}

Now we have all ingredients in hand to define the path integral of the regularized model. As we will see in the following sections, evaluation of the propagator then reduces to a simple combinatorial problem and applications of the theory of critical phenomena.

\subsection{Discrete solution: The transfer matrix}\label{sec:transfer}

In this section we evaluate the gravitational path integral of the two dimensional CDT model following the procedure explained above \cite{Ambjorn:1998xu}.

\begin{figure}[t]
\begin{center}
\includegraphics[width=4in]{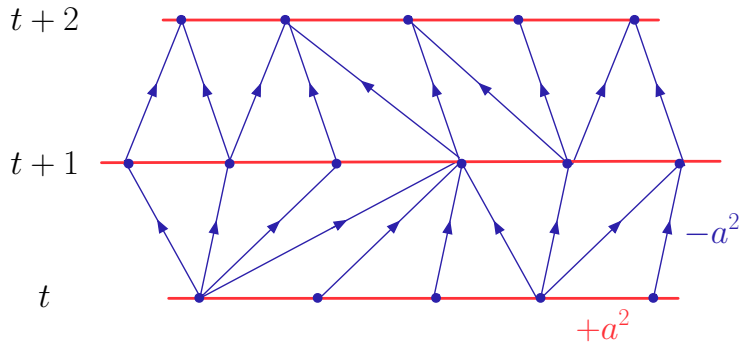}
\caption{Section of a 2d Lorentzian triangulation consisting of space-time strips of height $\Delta t\equ 1$. Each spatial slice is periodically identified, such that the simplicial manifold has topology $[0,1]\!\times\! S^1$. One sees that a single strip with lower boundary length $l_1$ and upper boundary length $l_2$ consists exactly of $l_1$ up pointing triangles and $l_2$ down pointing triangles.}
\label{fig:triangulation}
\end{center}
\end{figure}

The space-time manifolds we are considering consist of just one type of simplicial building blocks, namely, flat Minkowskian triangles with one space-like edge of length squared $l_s^2\equ a^2$ and two time-like edges of length squared $l_t^2\equ -a^2$. Each triangle then has a volume of $\frac{\sqrt{5}}{4}a^2$ (Appendix \ref{App:Lorentzian}). Our globally hyperbolic simplicial manifold is built up of space-time strips of height $\Delta t\equ 1$, where each strip consists of up and down pointing Minkowskian triangles (Figure \ref{fig:triangulation}). The foliation parameter $t$ is interpreted as the discretized version of ``proper'' time $T\equ a\cdot t$. Each spatial slice at time $t$ is chosen to have periodic boundary conditions with fixed  topology $S^1$ of the sphere. (Different boundary conditions are discussed in Appendix \ref{App:boundary}-\ref{App:hamilton}). Note that each spatial geometry is completely characterized by its discrete length $l$, where $L\equ a\cdot l$ is the spatial length. Further, due to the $S^1$ topology, the symmetry factor of a spatial slice of length $l$ is just $C(l)\equ l$.

After we have defined the triangulations contributing to the path integral, the next step is to determine their corresponding weights, i.e. the Regge action. Recall that the Einstein-Hilbert action in two dimensions is given by
\begin{equation}
S_{\mathrm{EH}}[g]=\int_M d^2x \sqrt{|\det g|}\, \left(\Lambda-K\,R(x)\right),
\end{equation}
where $K\equ G_N^{-1}$ is the inverse Newton's constant.
A simplification occurs due to the topological character of the curvature term in two dimension,
\begin{equation}\label{eq:trans:bonnet}
\int_M d^2x \sqrt{|\det g|}\, R(x)=2\pi \chi(M),
\end{equation}
where $\chi(M)\equ 2\mi 2\genus$ is the Euler characteristic of the manifold $M$ and $\genus$ is the genus of $M$. Hence, since we are not allowing for spatial topology changes, the exponential of $i$ times this term is just a constant phase factor which does not influence the dynamics and therefore can be pulled out of the path integral. If one actually allows for topology changes this term will be more involved as we will see in Section \ref{sec:Topo}. Since all triangles have the same volume, the Regge action \eqref{eq:simpl:reggeaction} takes the following simple form
\begin{equation}
\label{eq:trans:reggeaction}
S_{\mathrm{Regge}}(T)=\tilde{\lambda} \,a^2\, N(T),
\end{equation}
where $\tilde{\lambda}$ is the bare cosmological constant and $N(T)$ the number of triangles in the triangulation $T$. Note that a factor of $\frac{\sqrt{5}}{4}$ coming from the volume term has been absorbed into $\tilde{\lambda}$. Using \eqref{eq:simpl:pathsum} and \eqref{eq:trans:reggeaction}, the discrete gravitational path integral over the set of two dimensional causal simplicial manifolds $\Tri$ with an initial boundary $l_{in}$ and a final boundary $l_{out}$, consisting of $t\equ t_{out}-t_{in}$ time steps, can be written as
\begin{equation}\label{eq:trans:pathint}
G^{(1)}_{\tilde{\lambda}}(l_{in},l_{out};t)=\sum_{\substack{T\in \Tri: \\ l_{in}\rightarrow l_{out}}}\frac{1}{l_{in}}e^{i\,\tilde{\lambda}\,a^2\,N(T)}.
\end{equation}
For simplicity we can remove the symmetry factor $1/l_{in}$ by marking a vertex on the initial boundary, hence define
\begin{equation}
G_{\tilde{\lambda}}(l_{in},l_{out};t)=l_{in}\,G^{(1)}_{\tilde{\lambda}}(l_{in},l_{out};t).
\end{equation}
The unmarked propagator can be easily recovered at a later stage.

The next step of the calculation will be to Wick rotate $G_{\tilde{\lambda}}(l_{in},l_{out};t)$ to the Euclidean sector. Following the prescription for the Wick rotation $\mathcal{W}$ as defined in \eqref{eq:simpl:Wick}, we get
\begin{equation}
G_{\tilde{\lambda}}(l_{in},l_{out};t)=\sum_{\substack{T\in \Tri: \\ l_{in}\rightarrow l_{out}}}e^{i\,\tilde{\lambda}\,a^2\,N(T)} \xrightarrow[]{\mathcal{W}}
G_\lambda(l_{in},l_{out};t)=\sum_{\substack{T\in \mathcal{W}(\Tri): \\ l_{in}\rightarrow l_{out}}}e^{-\lambda\,a^2\,N(T)},
\end{equation}
where $\lambda$ and $\tilde{\lambda}$ differ by a $\mathcal{O}(1)$ constant, due to the volume difference of the Euclidean and Minkowskian triangles (Appendix \ref{App:Lorentzian}).\\

An essential quantity for the solution of the model is the \textit{transfer matrix} $\Trans$ which contains all dynamical information of the system. The transfer matrix can be represented by its matrix elements with respect to the initial state $\ket{l_{in}}$ and the final state $\ket{l_{out}}$, i.e. the kernel
\begin{equation}\label{eq:trans:transferematrix}
G_\lambda(l_{in},l_{out};t=1)=\bra{l_{out}}\Trans\ket{l_{in}},
\end{equation} 
which is often referred to as the \textit{one-step propagator}. From the semi-group property of the propagator,
\begin{eqnarray}
G_\lambda(l_{1},l_{2};t_{1}+t_{2})&=&\sum_{l} G_\lambda(l_{1},l;t_{1}) G_\lambda(l,l_{2};t_{2}),\\
G_\lambda(l_{1},l_{2};t_{1}+1)&=&\sum_{l} G_\lambda(l_{1},l;t_{1}) G_\lambda(l,l_{2};1),\label{eq:trans:semi}
\end{eqnarray}
it follows that the propagator $G_\lambda(l_{in},l_{out};t)$ for an arbitrary time $t$ is obtained by iterating \eqref{eq:trans:transferematrix} $t$ times, yielding
\begin{equation}
G_\lambda(l_{in},l_{out};t)=\bra{l_{in}}\Trans^t\ket{l_{out}}.
\end{equation}
In the continuum language this expression is closely related to the continuum partition function,
\begin{equation}
\mathcal{Z}_T= \lim_{\substack{a \rightarrow 0\\ t \rightarrow\infty}} \Trace \,\Trans^t,
\end{equation}
where in the limit the product $T\equ t\cdot a$ is kept fixed. 
One sees that knowing the eigenvalues of the transfer matrix is the key to solving the general problem.

Let us therefore concentrate on triangulations $T$ of just one single time step:
For such a strip with initial boundary of length $l_{in}$ and final boundary of length $l_{out}$ the total number of triangles is just $N(T)\equ l_{in}\pl l_{out}$ (cf. Figure \ref{fig:triangulation}). Evaluating the eigenvalues of the transfer matrix then reduces to a simple counting problem,
\begin{equation}
\label{eq:trans:onestepcomb}
G_\lambda(l_{in},l_{out};t=1)=e^{-\lambda\,a^2\,(l_{in}+l_{out})}\sum_{\substack{T\in \mathcal{W}(\Tri): \\ l_{in}\rightarrow l_{out}}} 1=e^{-\lambda\,a^2\,(l_{in}+l_{out})}\frac{l_{in}}{l_{in}+l_{out}}\binom{l_{in}+l_{out}}{l_{in}}.
\end{equation} 

For later purposes it is convenient to introduce the \textit{generating function} for the propagator $G_\lambda(l_{in},l_{out};t)$, defined as
\begin{equation}\label{eq:trans:generatingdef}
G(x,y;g;t)=\sum_{l_{in},l_{out}}G_\lambda(l_{in},l_{out};t)\,x^{l_{in}}y^{l_{out}},
\end{equation}
where we used the notation $g\equ e^{-\lambda\,a^2}$.
This expression can also be used to rewrite the semi-group property or composition law \eqref{eq:trans:semi}, yielding
\begin{equation}\label{eq:trans:semigen}
G(x,y;g;t_1+t_2)=\oint\frac{dz}{2\pi iz}\,G(x,z^{-1};g;t_1)G(z,y;g;t_2).
\end{equation}
The quantities $x$ and $y$ can be seen as purely technical tools of the generating function formalism, but one can also view them as boundary cosmological constants
\begin{equation}
x=e^{-\lambda_{in}\,a},\quad y=e^{-\lambda_{out}\,a}.
\end{equation}
This interpretation is useful in the context of renormalization as we will see in the next section. The generating function of the one-step propagator can be obtained from \eqref{eq:trans:onestepcomb} and \eqref{eq:trans:generatingdef} by usual techniques, yielding\footnote{Note that the asymmetry in $x$ and $y$ is due to the marking in the initial spatial boundary. ``Unmarking'' at a later stage will recover symmetry.}
\begin{equation}
\label{eq:trans:onestepgen}
G(x,y;g;1)=\frac{g^2xy}{(1-gx)(1-g(x+y))}
\end{equation}
The joint region of convergence of \eqref{eq:trans:onestepgen} is given by
\begin{equation}
\label{convergence }
|x|<1,\quad |y|<1,\quad |g|<\frac{1}{2}.
\end{equation}

Another way of obtaining \eqref{eq:trans:onestepgen} is by graphical methods. Associating a factor of $gx$ to every up pointing triangle ``$\bigtriangleup$'' and a factor of $gy$ to every down pointing triangle ``$\bigtriangledown$'', the sum over all possible triangulations of one strip with circular boundary conditions can be written as
\begin{eqnarray}
G(x,y;g;1) & = & \raisebox{-10pt}{\includegraphics[height=25pt]{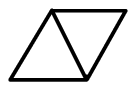}}+
\raisebox{-10pt}{\includegraphics[height=25pt]{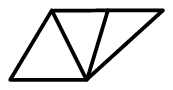}}+
\raisebox{-10pt}{\includegraphics[height=25pt]{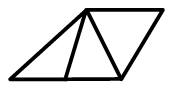}}+... \nonumber\\
 & = &  \sum_{k=0}^\infty\left(gx \sum_{l=0}^\infty (gy)^l\right)^k- \sum_{k=0}^\infty (gx)^k.\label{eq:trans:graph}
\end{eqnarray}
The subtraction of the last summand has been performed to remove the degenerate cases where either entrance or exit loop have length zero. Evaluation of \eqref{eq:trans:graph} shows the equivalence to \eqref{eq:trans:onestepgen}.

\subsection{Continuum limit}\label{sec:cont}

In this section we want to construct the continuum limit of the discretized model, by making use of standard techniques of the theory of critical phenomena. Note that all the calculations are performed in the Euclidean sector of the theory.

In terms of critical phenomena the \textit{continuum limit} is obtained by fine-tuning the coupling constants, generally denoted by $\lambda$, to the critical point $\lambda_c$ of the phase transition. This means a divergence of the \textit{correlation length} $\xi$ in the \textit{scaling limit}
\begin{equation}
\xi(\lambda)\sim\frac{1}{|\lambda-\lambda_c|^\nu}\xrightarrow[]{\lambda\rightarrow\lambda_c} \infty,
\end{equation}
while the cutoff $a$ goes to zero
\begin{equation}
a(\lambda)\sim\ |\lambda-\lambda_c|^\nu \xrightarrow[]{\lambda\rightarrow\lambda_c} 0,
\end{equation}
where $\nu\!>\!0$ is a critical exponent.
In terms of the cutoff $a$ and the correlation length $\xi$ the \textit{physical correlation length} is given by $L\equ \xi(\lambda)\cdot a(\lambda)$ and the continuum limit is defined as the simultaneous limit
\begin{equation}
a(\lambda)\rightarrow 0,\quad \xi(\lambda)\rightarrow \infty \quad\text{for fixed}\quad L=\xi(\lambda)\cdot a(\lambda).
\end{equation}
Having defined the general scheme to perform the continuum limit, let us now come back to the concrete problem of the discretized model described by the one-step propagator \eqref{eq:trans:onestepgen}.

From standard techniques of renormalization theory, we expect the couplings with positive mass dimension, i.e. the cosmological constant $\lambda$ and the boundary cosmological constants $\lambda_{in}$ and $\lambda_{out}$ to undergo an additive renormalization,
\begin{equation}
\lambda=\frac{A_\lambda}{a^2}+\Lambda,\quad
\lambda_{in}=\frac{A_{\lambda_{in}}}{a}+X,\quad
\lambda_{out}=\frac{A_{\lambda_{out}}}{a}+Y,
\end{equation}
where $\Lambda$, $X$ and $Y$ denote the corresponding renormalized values.
Introducing the critical values
\begin{equation}
g_c=e^{-A_\lambda},\quad x_c=e^{-A_{\lambda_{in}}},\quad y_c=e^{-A_{\lambda_{out}}},
\end{equation}
it follows that
\begin{equation}
g=g_c \,e^{-a^2\,\Lambda},\quad x=x_c \,e^{-a\,X},\quad y=y_c \,e^{-a\,Y}.
\end{equation}
The continuum limit can be performed by fine-tuning to the critical values $\mbox{$(x,y,g)\mapsto (x_c,y_c,g_c)$}$, as explained above. Further, it can be shown \cite{Ambjorn:1998xu}, that the only sensible continuum theory can be obtained by choosing $(x_c,y_c,g_c)\equ (1,1,\frac{1}{2})$ or $(x_c,y_c,g_c)\equ (-1,-1,-\frac{1}{2})$, which both lead to the same continuum theory. Hence, performing the continuum limit can be done by simultaneously fine-tuning to the critical values with use of the following scaling relations:
\begin{eqnarray}
g & = & \frac{1}{2}(1-a^2\,\Lambda) + \mathcal{O}(a^{3}), 
\label{eq:cont:scaling1}\\
x & = & 1-a\,X + \mathcal{O}(a^{2}),\label{eq:cont:scaling2}\\
y & = & 1-a\,Y + \mathcal{O}(a^{2})\label{eq:cont:scaling3}.
\end{eqnarray}

Let us now show how we can reveal the continuum Hamiltonian in this scaling limit. First note that, as we have rewritten the composition law in terms of generating functions, i.e. \eqref{eq:trans:semigen}, we can also rewrite the time evolution of the wave function (the analogue of \eqref{eq:path:timeevolpos}) in a similar manner,
\begin{equation}
\label{eq:cont:timeevol}
\psi(x,t+1)= \oint \frac{dz}{2 \pi i z} G(x,z^{-1};1)\psi(z,t).
\end{equation}
Introducing the scalings \eqref{eq:cont:scaling1}-\eqref{eq:cont:scaling3} and $t\equ \frac{T}{a}$ into \eqref{eq:cont:timeevol} and expanding both sides to order $a$ gives
\begin{eqnarray}
  \!\!&\!&\!\left(1-a \hat{H}(X,\partial_X)+\mathcal{O}(a^2)\right) \psi(X)  \nonumber\\ 
    &=&\int^{i\infty}_{-i\infty} \frac{dZ}{2\pi i} \left\{\left(\frac{1}{Z-X} +
    + a\,\frac{2\,X^2-4XZ+Z^2+2\,\Lambda}{(Z-X)^2}\right) + \mathcal{O}(a^2)\right\} \psi(Z),\label{eq:cont:expansion}
\end{eqnarray}
where we have defined $\psi(X)\!\equiv\!\psi(x\equ 1-a\, X)$ and respectively for $Z$. For the expansion of the left hand side of \eqref{eq:cont:expansion} we used the formula for an infinitesimal time evolution of the wave function which is also used in the fundamental definition of the path integral,
\begin{equation}
\psi(X,T+a)=e^{-a\,\hat{H}(X,\partial_X)}\psi(X,T).
\end{equation}
Notice that the first term on the right hand side of \eqref{eq:cont:timeevol}, $\frac{1}{Z-X}$, is the inverse Laplace transformed delta function $\delta(Z-X)$ as expected. Performing the integration in \eqref{eq:cont:expansion}, one can extract the Hamiltonian
\begin{equation}
\label{eq:cont:HamX}
\hat{H}(X,\deriv{}{X})=X^2\,\deriv{}{X}+2\,X-2\,\Lambda\deriv{}{X}.
\end{equation}
It is important to notice at this point that the Hamiltonian \eqref{eq:cont:HamX} does not depend on possible higher order terms of the scaling relations  \eqref{eq:cont:scaling1}-\eqref{eq:cont:scaling3}. One might check this by explicitly introducing higher order terms in the scaling relations, hence
\begin{eqnarray}
x &=& 1-a\,X +\frac{1}{2}\gamma\,a^2\,X^2+\mathcal{O}(a^3),\label{eq:cont:scaling2new}\\
y &= & 1-a\,Y
+\frac{1}{2}\gamma\,a^2\,Y^2+\mathcal{O}(a^3)\label{eq:cont:scaling3new}.
\end{eqnarray}
Following the above procedure with the new scaling relations \eqref{eq:cont:scaling2new}-\eqref{eq:cont:scaling3new}, one sees that the resulting Hamiltonian does not depend on $\gamma$ and is still described by \eqref{eq:cont:HamX}.

Having obtained the effective quantum Hamiltonian in ``momentum''-space $X$, one can go back to the Hamiltonian in length-space $L$ by performing an inverse Laplace transformation on the wave functions,
\begin{equation}
\psi(L)=\int^{i\infty}_{-i\infty} \frac{dX}{2\pi i} \,e^{XL}\psi(X).
\end{equation}
The inverse Laplace transformation is the continuum counterpart of the defining equation for the generating functions, formula \eqref{eq:trans:generatingdef}, which one can think of as a discrete Laplace transformation. The \textit{effective quantum Hamiltonian} for the marked propagator then reads
\begin{equation}
\label{eq:cont:HamLmarked}
\hat{H}(L,\dL)=-L\ddL+2\Lambda L,\quad\text{(marked)}.
\end{equation}
This operator is selfadjoint on the Hilbert space $\mathcal{H}\equ \mathcal{L}^2(\mathbb{R}_+,L^{-1}dL)$. The effective quantum Hamiltonian corresponding to the unmarked propagator can then be obtained as
\begin{equation}
\label{eq:cont:HamL}
\hat{H}(L,\dL)=-L\ddL-2\dL+2\Lambda L,\quad\text{(unmarked)},
\end{equation}
which is selfadjoint on the Hilbert space $\mathcal{H}\equ \mathcal{L}^2(\mathbb{R}_+,LdL)$. Notice that both Hamilton operators describe the same physical system.\footnote{Consider the symmetric matrix element for the Hamiltonian of the marked case $\mbox{$L_1^{-1}dL_1\bra{L_1}\hat{H}\ket{L_1}$}$. From \eqref{eq:cont:propagator} it follows that unmarking can be done by absorbing a factor $L_1^{-1}$ into the wave function, i.e. $\ket{L_1}\mapsto\ket{L_1^u}\equ L^{-1}\ket{L_1}$.  The unmarked Hamiltonian can then be obtained by evaluating $\mbox{$dL_1^{-1}dL_1\bra{L_1}\hat{H}\ket{L_1}\equ L_1dL_1\bra{L_1^u}\hat{H}^u\ket{L_1^u}$}$.}
In the following we want to analyze the effective quantum Hamiltonian \eqref{eq:cont:HamL}, calculate the spectrum, the eigenfunctions and from this further quantities like the partition function and the finite time propagator. A detailed calculation of these results is presented in Appendix \ref{App:Calogero}.

The Hamiltonian consists of a kinetic term which depends on the spatial length of the ``universe'' and a potential term which depends on the (renormalized) cosmological constant. Further, it is important to notice that the Hamiltonian is bounded from below and therefore leads to a well-defined quantum theory. The spectrum of the Hamiltonian is discrete with equidistant eigenvalues,
\begin{equation}
\label{eq:cont:Energy}
E_n=2\sqrt{2 \Lambda}(n+1)\quad n=0,1,2,..
\end{equation}
The corresponding eigenfunctions are given by
\begin{equation}
\label{eq:cont:eigenfunc}
\psi_n(L)=\mathcal{A}_n e^{-\sqrt{2\Lambda}L}\, {}_1F_{1}(-n;2;2\sqrt{2\Lambda}L),\quad d\mu(L)=LdL,
\end{equation}
where ${}_1F_{1}(-n,\mu+1;z)$ are the confluent hypergeometric functions. In this case the power series of the confluent hypergeometric function is truncated to a polynomial of degree $n$, namely, the generalized Laguerre polynomials, here denoted by $L_n^\mu(z)$,
\begin{eqnarray}
{}_1F_{1}(-n;\mu+1;z)&=&\frac{\Gamma(n+1)\Gamma(\mu+1)}{\Gamma(n+\mu+1)}\,L_n^\mu(z)\\
&=&\sum_{k=0}^n (-1)^k \binom{n}{k}\frac{1}{(\mu+1)(\mu+2)...(\mu+k)}\frac{z^k}{k!}.\label{eq:cont:hypergeom}
\end{eqnarray}
From this one can see that the eigenfunctions $\psi_n(L)$ form an orthonormal basis of the Hilbert space $\mathcal{H}$, where the normalization factor is given by
\begin{equation}
\mathcal{A}_n=2\sqrt{2\Lambda(n+1)}.
\end{equation}
Having obtained the spectrum of the Hamiltonian one can easily calculate the (Euclidean) \textit{partition function} by using
\begin{equation}
\label{eq:cont:partitionfunc}
\mathcal{Z}_T=\Trace (e^{-T\hat{H}})=\sum_{n=0}^\infty e^{-TE_n}.
\end{equation}
Hence, inserting \eqref{eq:cont:Energy} into \eqref{eq:cont:partitionfunc} yields
\begin{equation}
\label{eq:cont:partitionfuncres}
\mathcal{Z}_T=\frac{1}{e^{2\sqrt{2\Lambda} T} -1}=\frac{1}{2}\left(\coth(\sqrt{2\Lambda}T)-1 \right).
\end{equation}
Another interesting quantity to calculate is the finite time propagator or loop-loop correlator which is defined by
\begin{eqnarray}
    G_\Lambda(L_1,L_2;T)&=&\bra{L_2}e^{-T\hat{H}}\ket{L_1}\\
    &=& \sum_{n=0}^{\infty}e^{-TE_n}\psi_n^*(L_2)\psi_n(L_1)\label{eq:cont:propagator},
\end{eqnarray}
where in the second line we used that the eigenstates form a complete orthonormal basis of the Hilbert space. Upon inserting \eqref{eq:cont:eigenfunc} into \eqref{eq:cont:propagator} one obtains the unmarked finite time continuum propagator (cf. Appendix \ref{App:Calogero})
\begin{eqnarray}
	G_\Lambda(L_1,L_2;T)=\sqrt{\frac{2\Lambda}{L_1L_2}}
	\frac{e^{-\sqrt{2 \Lambda}(L_1+L_2)\coth(\sqrt{2\Lambda}T)}}{\sinh(\sqrt{2\Lambda}T)}
	I_1 \left(2 \frac{\sqrt{2 \Lambda L_1 L_2}}{\sinh(\sqrt{2\Lambda}T)}\right)\label{eq:cont:propagatorres},
\end{eqnarray}
where $I_1(x)$ denotes the modified Bessel function of the first kind. The corresponding propagator with a marking on the initial (final) spatial boundary can then be obtained by multiplying \eqref{eq:cont:propagatorres} with $L_1$ ($L_2$). The finite time propagator \eqref{eq:cont:propagatorres} of two-dimensional CDT was first obtained in \cite{Ambjorn:1998xu}.

A remarkable fact to mention at this point is that, up to some marking ambiguities, the finite time propagator \eqref{eq:cont:propagatorres} agrees with the result of the propagator obtained from a continuum calculation in proper-time gauge of two-dimensional pure gravity \cite{Nakayama:1993we}.

Since all continuum results presented so far were calculated in the Euclidean sector, the remaining task is to define an inverse Wick rotation to relate the Euclidean sector with the Lorentzian sector of the theory. A natural proposal is to analytically continue the continuum proper time $T\mapsto -iT$. Under this prescription the effective quantum Hamiltonian corresponds to a unitary time evolution,
\begin{equation}
\U(T,T')=e^{i\hat{H}(T-T')},\quad \text{with}\quad \hat{H}(L,\dL)=-L\ddL-2\dL+2\Lambda L.
\end{equation}
Further, the Euclidean partition function \eqref{eq:cont:partitionfuncres} and the Euclidean finite time propagator \eqref{eq:cont:propagatorres} can then be simply related to their corresponding Lorentzian expressions. For the case of the finite time propagator, the expression in the Lorentzian sector reads,
\begin{equation}
G_\Lambda(L_1,L_2;T)=\sqrt{\frac{2\Lambda}{L_1L_2}}
	\frac{e^{i \sqrt{2 \Lambda}(L_1+L_2)\cot(\sqrt{2\Lambda}T)}}{i\sin(\sqrt{2\Lambda}T)}
	I_1 \left(2\frac{\sqrt{2 \Lambda L_1 L_2}}{i\sin(\sqrt{2\Lambda}T)}\right)\quad\text{(Lorentzian s.)}.
\end{equation}
Here one can clearly see that the propagator possesses oscillation modes in the time variable $T$.

As a final remark, notice that the definition of the inverse Wick rotation is not a priori clear. Besides the definition given above, one might also consider the analytic continuation in the cosmological constant $\Lambda\mapsto -i\Lambda$, which would also correspond to a continuation $a_t\mapsto -i a_t$ on the time-like cutoff. Unfortunately, it does not lead to a unitary theory, for which reason we prefer the analytic continuation on the proper time variable $T\mapsto -iT$ as the physically sensible inverse Wick rotation.    

\subsection{Physical observables} \label{sec:obs}

Having calculated the effective quantum Hamiltonian, the partition function and the propagator, there remains the question: What is the physics behind those expressions?
As one can see from \eqref{eq:trans:bonnet}, the classical Einstein equations in two dimensions are empty. Nevertheless, we obtained a non-trivial expression for the effective quantum Hamiltonian. Therefore, whatever dynamics it describes will be purely quantum, without a physical non-trivial classical limit. More precisely, we will see that it corresponds to quantum fluctuations of the spatial length of the ``universe''.
 
\begin{figure}[h]
\begin{center}
\includegraphics[width=2.5in]{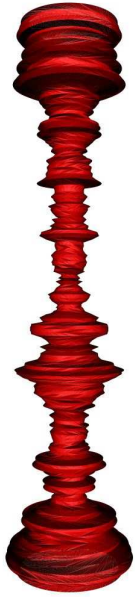}
\caption{A typical two-dimensional Lorentzian space-time. The compactified direction shows the spatial hypersurfaces of length $\expec{L}$ and the vertical axis labels proper time $T$. Technically, the picture was generated by a Monte-Carlo simulation, where a total volume of $N=18816$ triangles and a total proper time of $t=168$ steps was used. Further, initial and final boundary has been identified.}
\label{fig:universe}
\end{center}
\end{figure}

Interesting observables to calculate are the expectation values of the spatial length and higher moments. Since the only dimensionful constant in the model is the cosmological constant with dimension $[\Lambda]\equ \text{length}^{-2}$, all dimensionful quantities will appear in appropriate units of $\Lambda$. Using the expression of the wave function \eqref{eq:cont:eigenfunc}, one can calculate expectation values of spatial length and all moments
\begin{equation}\label{eq:obs:explength}
\expec{L^m}_n\equiv\bra{n}L^m\ket{n}=\int_0^\infty d\mu(L) \psi^*_n(L)\,L^m\,\psi_n(L). 
\end{equation}
A general expression for \eqref{eq:obs:explength} is presented in Appendix \ref{App:Calogero}. Let us just state the first two expressions here, namely,
\begin{equation}
\label{eq:obs:length}
\expec{L}_n=\frac{n+1}{\sqrt{2\Lambda}},\quad\expec{L^2}_n=\frac{3}{4}\frac{(n+1)^2}{\Lambda}.
\end{equation}
Generally, the moments scale as $\expec{L^m}_n\!\sim\! \Lambda^{-\frac{m}{2}}$ as shown in the appendix. From \eqref{eq:obs:length} one can also calculate the variance $\expec{\Delta L}_n$ of the spatial length,
\begin{equation}
\expec{\Delta L}_n=\sqrt{\expec{L^2}_n-\expec{L}_n^2}=\frac{1}{2}\,\frac{n+1}{\sqrt\Lambda}.
\end{equation}
One can see that for the ``universe'' in a certain state $\ket{n}$, for example, the ground state $\ket{0}$, it possesses fluctuations around the average value $\expec{L}_0$ with variance $\expec{\Delta L}_0$. This behavior is illustrated in Figure \ref{fig:universe} by a numerical Monte-Carlo simulation. What is also good to see in the figure is that the fluctuations are roughly of the same order as the spatial length of the ``universe'', i.e. $\expec{\Delta L}\!\sim\!\expec{L}$.

Other useful quantities to calculate are certain critical exponents of the continuum theory, which, due to the gravitational setting, will all have a geometrical interpretation. Among those critical exponents, there is one of special interest, namely, the \textit{Hausdorff dimension} $d_H$ (cf. \cite{Ambjorn:1997di}). In particular, $d_H$ can be calculated for each geometry $[g_{\mu\nu}]$, but what we will consider here is the ensemble average over the entire class $\Geom(M)$, that is, the leading order scaling behavior of the expectation value
\begin{equation}
\label{eq:ref:hausdorff}
\expec{V}_T\sim T^{d_H}.
\end{equation}
To calculate this quantity we use the partition function \eqref{eq:cont:partitionfuncres}, whose behavior for large $T$ is given by
\begin{equation}
\label{eq:obs:ZlargeT}
\mathcal{Z}_T(\Lambda)\sim e^{-2\sqrt{2\Lambda}T}\quad\text{for}\quad T\rightarrow\infty.
\end{equation} 
From this we can define a typical energy scale $M(\Lambda)$ of the ``universe'',
\begin{equation}
M(\Lambda)=\lim_{T\rightarrow\infty}\left( -\frac{\log \mathcal{Z}_T(\Lambda)}{T} \right)\sim\Lambda^{\frac{1}{2}}.
\end{equation}
On the other hand, we can use the partition function $\mathcal{Z}_T(\Lambda)$ to calculate the ensemble average of the volume, by taking derivatives with respect to the corresponding coupling, namely, the cosmological constant $\Lambda$,
\begin{equation}
\label{eq:obs:VfromZ}
\expec{V}_T=-\frac{1}{\mathcal{Z}_T(\Lambda)}\deriv{\mathcal{Z}_T(\Lambda)}{\Lambda}.
\end{equation}
Upon inserting \eqref{eq:obs:ZlargeT} into \eqref{eq:obs:VfromZ}, we can determine the scaling behavior of the volume at large scales $T \!\gg\! 1/M(\Lambda)$,
\begin{equation}\label{eq:obs:largeTV}
\expec{V}_T\sim T\cdot \Lambda^{-\frac{1}{2}}\quad \text{for}\quad T\gg\frac{1}{M(\Lambda)}.
\end{equation}
This shows that at large scales $T\!\gg\! 1/M(\Lambda)$, the typical universe has a volume which is proportional to $T$ and therefore looks like a long tube of length $T$. Further, from \eqref{eq:obs:largeTV} one can read off that the spatial length scales as
\begin{equation}
\expec{L}=\frac{\expec{V}}{T}\sim\frac{1}{\sqrt{\Lambda}},
\end{equation}
as we have already found in \eqref{eq:obs:length}. More interesting than the scaling of the volume at large scales is the typical scaling at small scales, namely, those of order $T\!\sim\! 1/M(\Lambda)$. In the same procedure one obtains
\begin{equation}
\expec{V}_T\sim T^2 \quad \text{for}\quad T\sim\frac{1}{M(\Lambda)}.
\end{equation}
From definition \eqref{eq:ref:hausdorff}, one then sees that the Hausdorff dimension of 2d Lorentzian quantum gravity or CDT is given by
\begin{equation}
d_H=2\quad\text{(CDT)}.
\end{equation}
Naively this looks trivial, since we were considering two-dimensional CDT which in the construction uses two-dimensional building blocks. However, the Hausdorff dimension is truly a dynamical quantity and is a priori \textit{not} the same as the dimension of the simplicial building blocks. An example of this is two-dimensional Euclidean quantum gravity defined through dynamical triangulations which has a Hausdorff dimension of $d_H\equ 4$.\footnote{One can rapidly check this by noting that the partition function in this case scales as $\mathcal{Z}_T(\Lambda)\sim e^{-c\sqrt[4]{\Lambda}T}$ for $T\rightarrow\infty$ \cite{Ambjorn:1998xu}.} This non-canonical value of the Hausdorff dimension is due to the highly degenerate, acausal geometries contributing to the ensemble average. At this point one can clearly see, as already mentioned in Section \ref{sec:lorentziannature}, that the continuum theories of two-dimensional quantum gravity with Euclidean and Lorentzian signature are distinct and cannot be related in a trivial manner (cf. footnote \ref{foot:EuLor})! \\

To conclude, we have seen that CDT gives a powerful method to define a nonperturbative path integral for quantum gravity. In the considered two-dimensional model with fixed cylindrical topology, it led to a well-defined continuum quantum theory, whose time evolution was unitary. Further, we have seen that the restriction to the causal geometries in the path integral led to a physically interesting ground state of a quantum ``universe'', which also showed that quantum gravity with Euclidean and Lorentzian signature are distinct theories.

One can now ask the question whether or not causal constraints might be a helpful tool to implement topology changes and the sum over topologies in the path integral. Finding a detailed answer to this question will be the topic of the following Section \ref{sec:Topo}.

Another aspect we have not considered so far is the question of universality of the theory: In general one would expect that the continuum theory should be independent of the details of the discretization procedure we use, e.g., whether we build up our space-time from triangles, squares or $p$-polygons should not affect any continuum physical observable. A discussion of this point will be the topic of Section \ref{sec:univers}.

\section{2D Lorentzian quantum gravity with topology change} \label{sec:Topo}

In this section we construct a combined path integral over geometries and topologies for two-dimensional Lorentzian quantum gravity. The section is structured as follows: In Section \ref{sec:sumtop} we introduce the reader to the general concepts of a path integral formulation for quantum gravity including a sum over topologies and discuss certain problems, especially in the context of Euclidean formulations. In Section \ref{sec:implemholes} we demonstrate qualitatively how the Lorentzian structure can be used to exclude certain geometries from the path integral  which lead to macroscopic causality violation; this makes the path integral well-behaved. In Section \ref{sec:toptrans} an explicit realization of this construction in the framework of two-dimensional CDT and the discrete solution thereof is presented. The difficulty in the construction of the continuum limit is to find a suitable double scaling limit for both the cosmological and Newton's constant, which we address in Section \ref{sec:topocont}. The resulting continuum theory of quantum gravity describes a quantum ``universe'' with fluctuating geometry and topology, as we will discuss in Section \ref{sec:topobserv} in terms of its physical observables. Further, we show in Section \ref{sec:topotaming} that the presence of infinitesimal wormholes in space-time leads to a decrease in the effective cosmological constant.

\subsection{Sum over topologies in quantum gravity} \label{sec:sumtop}

In Section \ref{sec:CDT} we have given a concrete meaning to the nonperturbative gravitational path integral as an integral over all possible geometries $\Geom(M)$,
\begin{equation}
\label{eq:sumover:Zfixed}
\mathcal{Z}=\int_{\Geom(M)} \D[g_{\mu\nu}] e^{i S_{\mathrm{EH}}[g_{\mu\nu}]},
\end{equation}
which by virtue of the approach of causal dynamical triangulations led to a well defined continuum theory of quantum gravity in two dimensions.

Nevertheless, many other attempts of constructing a nonperturbative gravitational path integral start from the ansatz that one should not only consider all geometries of a manifold $M$ of a certain fixed topology, but further include the \textit{sum over all space-time topologies} in the gravitational path integral, formally written as
\begin{equation}
\label{eq:sumover:Z}
\mathcal{Z}=\sum_M\int_{\Geom(M)} \D[g_{\mu\nu}] e^{i S_{\mathrm{EH}}[g_{\mu\nu}]}.
\end{equation}  
But how should we think about summing over topologies and further about the associated \textit{topology changes} \cite{Dowker:2002hm} of space as a function of time? Clearly, theories which predict topology changes at \textit{macroscopic} scales are unlikely to be consistent with observational data. A suggestion for how one should think about topology changes is therefore as topological excitations of space-time at very short scales such as the Planck scale. This leads to the notion of \textit{space-time foam} \cite{Wheeler,Garay:1999cy}, according to which space-time is a smooth manifold at macroscopic scales, but at small scales is dominated by roughly fluctuating geometries and topologies. Whether nature really admits this behavior and one should include the sum over topologies into the path integral is still unclear.

The gravitational path integral including the sum over topologies has been the subject of intensive study in the context of four-dimensional \textit{Euclidean} quantum gravity \cite{Hawking:1979zw,Hawking:1978jz}. However, in nonperturbative quantum gravity models, where one can analyze the problem explicitly, it turns out that topology changing configurations completely dominate the path integral \eqref{eq:sumover:Z}, since the number of contributing geometries grows \textit{super-exponentially} with the volume $N$, i.e. at least $\sim\! N!$, which makes the path integral badly divergent.

Moreover, this problem is still present in Euclidean quantum gravity in dimensions $d\kl 4$. In the case of two dimensions, Euclidean quantum gravity without topology changes is well understood analytically, as mentioned in Section \ref{sec:CDT} in the context of dynamical triangulations. Furthermore, the sum over topologies takes a very simple form, as a sum over a single parameter, $\genus \grgl 0$, the genus of the space-time manifold $M$, i.e. the number of space-time handles. Including the sum over topologies, the Euclidean analogue of \eqref{eq:sumover:Z} has been studied intensively, also because it can be interpreted as a nonperturbative sum over worldsheets of a bosonic string with zero-dimensional target space \cite{Douglas:1989ve}. There it turns out that the divergence of the path integral still persists, moreover, the topological expansion of \eqref{eq:sumover:Z} is not even Borel-summable \cite{DiFrancesco:1993nw}. Further attempts to solve this problem in the Euclidean context have so far been unsuccessful.

To draw a lesson from Section \ref{sec:CDT}, we have seen that in pure Lorentzian quantum gravity or CDT (we will from now on refer to the CDT model with fixed topology as the ``pure'' model) the restriction to causal geometries in the gravitational path integral \eqref{eq:sumover:Zfixed} leads to a physically interesting ground state of a quantum ``universe'' which is distinct from the Euclidean case. Inspired by this, one might now ask the question whether implementing the (almost everywhere) causal structure into the gravitational path integral including the sum over topologies might restrict the state space in such a way that the number of geometries does not grow super-exponentially anymore. That this is indeed the case, and that one can unambiguously define \eqref{eq:sumover:Z} in the setting of CDT has been shown in \cite{Loll:2003rn,Loll:2005dr}\footnote{Also see \cite{Loll:2003yu,conf} for additional information.} and will be discussed in the following sections.

\subsection{Implementing topology changes}\label{sec:implemholes}

Having adopted the notion of space-time foam and the sum over topologies in the path integral, we have to face the question, what kind of topology changing geometries should contribute to the sum in \eqref{eq:sumover:Z}. 

From observational data we tend to exclude all topology changing geometries which lead directly to a \textit{macroscopic} acausal behavior. In our implementation we therefore consider only topology changes which are of an infinitesimal duration, so-called \textit{infinitesimal wormholes}. In terms of the framework of CDT this means that wormholes do only exist within one time step $\Delta t\equ 1$ of discrete proper time. Further, the total number of wormholes per such discrete time step can be arbitrary in the continuum limit. Using this notion of infinitesimal wormholes, we consider triangulations whose spatial slices have topology $S^1$ for integer values of $t$, whereas in the interval $]t,t\pl 1[$ it splits into $\genus_t\pl 1$ $S^1$-components, where $\genus_t$ is the number of wormholes or equivalently the genus of this space-time strip. 

The setting with infinitesimal wormholes in a path integral formulation of quantum gravity with a sum over topologies has been studied before in the context of Euclidean quantum gravity in four dimensions \cite{Coleman:1988tj,Klebanov:1988eh}. Nevertheless, as we will show in the following, even generic space-times adopting this mildest form of topology change can cause a super-exponential growth in the number of configurations in the path integral or lead to a macroscopic acausal structure. The essential difference between Euclidean quantum gravity and CDT is the (almost everywhere) Lorentzian structure, which enables us to classify how badly causality is violated and to exclude certain ``badly'' causality violating geometries from the path integral. To be more precise on this point, clearly the metric of the topology changing geometries becomes degenerate at the saddle points $p$, where space-time splits or merges. Nevertheless, one is still able to assign a light-cone to these so-called Morse points by analysis of the neighboring points as discussed in \cite{Dowker:1997hj,Borde:1999md,Dowker:1999wu,Dowker:1999cp}. 

Let us now give some qualitative arguments for excluding certain classes of geometries with ``bad'' topology changes, and then present the concrete realization of how to implement the remaining ``not-so-bad'' topology changes in the framework of CDT in the next section.

\begin{figure}[t]
\begin{center}
\includegraphics[width=6in]{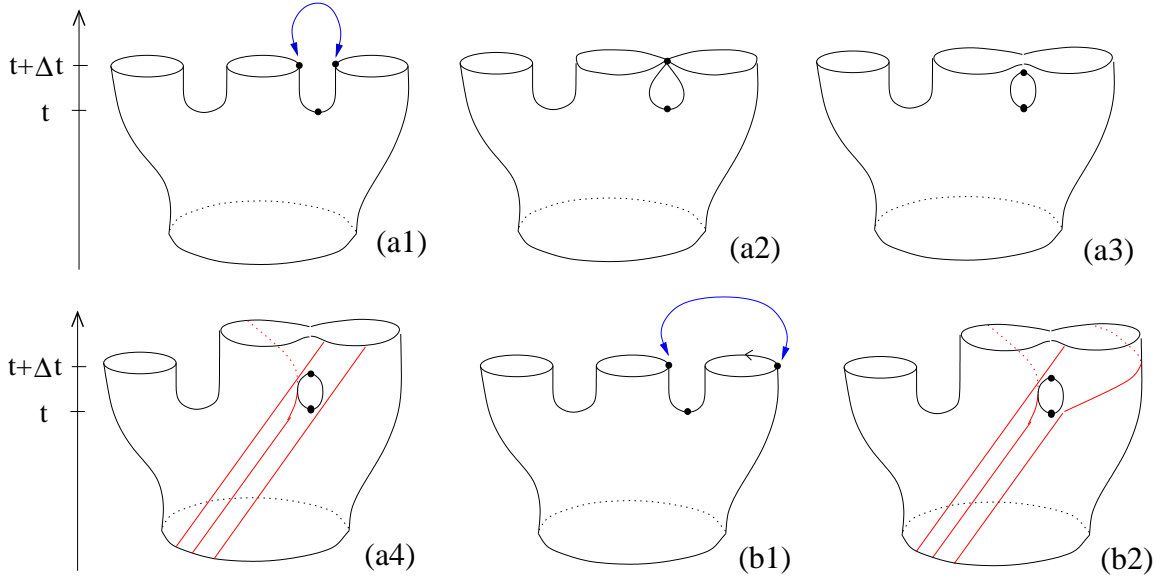}
\caption{(a1) A spatial slice of topology $S^1$ at time $t$ splits into three components giving rise to two saddle points $p_t$. (a2) A merging of two components at time $t+1$, where the saddle points $p_t$ and $p_{t+1}$ are time-like related. This gives rise to a ``untwisted'' wormhole, as shown in (a3). Parallel light rays passing this wormhole are unaffected unless they are not scattered directly by the wormhole, as emphasized in (a4). For the case of a twisted regluing (b1), where the saddle points $p_t$ and $p_{t+1}$ are space-like separated, parallel light beams passing this wormhole split into two parts with a relative separation (b2).}
\label{fig:tubes}
\end{center}
\end{figure}

The creation of wormholes is illustrated in Figure \ref{fig:tubes}. At discrete proper time $t$ the spatial slice of topology $S^1$ splits into $\genus_t\pl 1$ $S^1$-components giving rise to $\genus_t$ saddle points (a1). After a time lapse $\Delta t \equ 1$, the $S^1$-components merge again to a single $S^1$,
which leads to $\genus_t$ wormholes in the space-time strip $[t,t\pl 1]$. An important point in this construction is the question of how one is going to reglue the different $S^1$-components at time $t+1$. It is not difficult to see that arbitrary regluings at time $t\pl 1$ give again rise to a super-exponential growth of the number of configurations. However, most of these regluings are very ill-behaved with respect to their ``causal structure'', in the sense that part of light beams can get a global rearrangement when passing through the wormholes. Fortunately, the imposed (almost everywhere) Lorentzian structure enables us to restrict ourselves to a subclass of topology changes which does not admit this behavior, namely, regluings without a relative rearrangement or twist of the corresponding $S^1$-components. These topology changes are characterized by the fact that the upper saddle point $p_{t+1}$ is time- or light-like related to the lower saddle point $p_t$, as indicated in Figure \ref{fig:tubes}, (a2)-(a3), where for simplicity only one merging of two $S^1$-components is shown.  

To illustrate the qualitative difference between those untwisted and twisted topology changes, consider light beams propagating through the resulting space-times, as shown in Figure \ref{fig:tubes}, (a4) and (b2). In both cases light beams which hit the wormhole scatter non-trivially. However, in the twisted case another effect takes place, namely, due to the relative spatial shift $\Delta l$ between the saddle points $p_t$ and $p_{t+1}$, parallel light beams passing this wormhole will split into two parts with a relative separation $\Delta l$. While the effect due to the non-trivial light scattering goes to zero in the continuum limit, the relative separation in the twisted case persists, even after the wormhole has disappeared. We therefore discard these twisted mergings as ``bad'' topology changes and exclude them from the path integral. Thus, we only consider geometries with untwisted wormholes in the sum over topologies. Note that even though the effect of light scattering at such an untwisted wormhole goes to zero in the continuum limit, this does not mean that the contribution of several wormholes will not lead to a measurable effect; in fact, this is precisely what we will observe in the resulting theory. 

\subsection{Discrete solution: The transfer matrix}\label{sec:toptrans}

In this section we discuss how wormholes of the type explained above can be introduced in the setting of CDT. Further, we present the combinatorial solution of the one-step propagator including the sum over topologies in the discrete setting.

\begin{figure}[t]
\begin{center}
\includegraphics[width=3.5in]{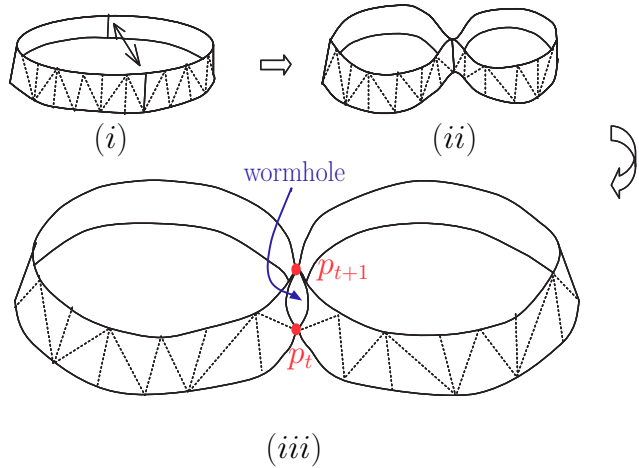}
\caption{Construction of a wormhole: starting from a space-time strip of topology $\mbox{$[0,1]\!\times\! S^1$}$ as in the pure CDT model (i), one identifies two time-like edged (ii) and then cuts open the geometry perpendicular to this line (iii). The two resulting saddle points at time $t$ and $t\pl 1$ are labeled with $p_t$ and $p_{t+1}$.}
\label{fig:makinghole}
\end{center}
\end{figure}

\begin{figure}[t]
\begin{center}
\includegraphics[width=3in]{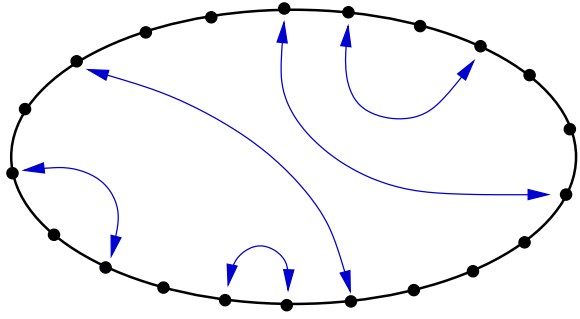}
\caption{Identification of time-like edges: Displayed is a cut through a space-time of topology $[0,1]\!\times\! S^1$ halfway between time $t$ and $t\pl 1$. The $N\equ l_{in}\pl l_{out}$ time-like edges appear as dots on a circle in the cutting plane. Pairwise identification of the $2\genus$ time-like edges (dots) is illustrated by the non-intersecting arches, where maximally one arch can connect to each dot.}
\label{fig:arch}
\end{center}
\end{figure}

Consider a typical space-time strip of topology $[0,1]\!\times\! S^1$, as used in the discrete setting of the pure CDT model, which is illustrated in Figure \ref{fig:makinghole}, (i). An ``untwisted'' wormhole of infinitesimal duration $\Delta t\equ 1$, as associated to a ``not-so-bad'' geometry, can be created by identifying two time-like edges of this space-time strip (ii) and cutting open the geometry perpendicular to this line (iii). Clearly, the resulting saddle points $p_t$ and $p_{t+1}$ are time-like related. The curvature singularities at those saddle points will be of the standard conical type after we have performed the Wick rotation and we can assign the Boltzmann weight accordingly.\footnote{See \cite{Louko:1995jw} for a related discussion.} A space-time strip of arbitrary genus $\genus$ can then be generated by repeating this procedure $\genus$-times, where the $\genus$ arrows identifying time-like edges are not allowed to intersect, as illustrated in Figure \ref{fig:arch}. This type of identification ensures that we obtain just an exponential growth in the volume, which leads to a well-defined continuum theory, in contrast to the super-exponential growth in the Euclidean model. 

As in the pure CDT model, all dynamical information is encoded in the transfer matrix and it therefore suffices to investigate the combinatorics of a single space-time strip. Hence, following the prescriptions of Section \ref{sec:transfer} and including the topological term \eqref{eq:trans:bonnet} in the Regge action, we obtain the Wick rotated one-step propagator 
\begin{equation}\label{eq:topotrans:sumovertop}
G_{\lambda,\kappa}(l_{in},l_{out};1)=e^{-\lambda\,a^2\,(l_{in}+l_{out})} \sum_{\substack{T\in \mathcal{W}(\Tri): \\ l_{in}\rightarrow l_{out}}}e^{-2\kappa\, \genus (T)},
\end{equation}
where the sum is taken over all possible triangulations $T$ of height $\Delta t\equ 1$ with fixed initial boundary $l_{in}$ and final boundary $l_{out}$, but arbitrary genus $0\klgl\genus (T)\klgl \left[N/2\right]$ (here $N\equ l_{in}\pl l_{out}$ is the number of triangles in $T$, which coincides with the number of time-like edges). Further, $\kappa$ is the bare inverse Newton's constant and $\lambda$ the bare cosmological constant. As in the pure model we have omitted an overall constant phase factor. 

Let us now perform the sum in \eqref{eq:topotrans:sumovertop} for fixed genus. Consider first the case without wormholes. To simplify the combinatorial expressions, we will use the combinatorial factor belonging to a space-time strip with open boundary conditions, instead of circular boundary conditions. As shown in Appendix \ref{App:boundary}-\ref{App:hamilton}, the resulting continuum theories in the pure CDT model are very similar for both boundary conditions, which enables us to recover the result for circular boundary conditions at a later stage. Hence, the combinatorial factor assigned to the sum over all possible triangulations without wormholes reads (Appendix \ref{App:boundary})
\begin{equation}
\tilde{G}(l_{in},l_{out})=\binom{l_{in}+l_{out}}{l_{in}}.
\end{equation}   
For a given strip with fixed number of triangles $N\equ l_{in}\pl l_{out}$, wormholes are created by the procedure explained above. Thereby, to construct a space-time strip with $0\klgl\genus (T)\klgl \left[N/2\right]$ wormholes, one first has to choose $2\genus$ out of $N$ time-like edges; the combinatorial factor assigned to this is simply given by
\begin{equation}
\binom{N}{2\genus}.
\end{equation}
For a given set of $2 \genus$ time-like edges we have to count the number of possibilities to pairwise identify these edges, where the arches belonging to the identifications are not allowed to intersect, as illustrated in Figure \ref{fig:arch}. This is a well-known combinatorial problem whose solution is given by the Catalan numbers
\begin{equation}
c_{\genus}=\frac{(2\genus)!}{\genus!(\genus+1)!}=\frac{1}{\genus+1}\binom{2\genus}{\genus}.
\end{equation}
Thus, the complete formula for the one-step propagator \eqref{eq:topotrans:sumovertop}, after the fixed genus part of the sum has been performed, reads
\begin{equation}\label{eq:toptrans:sumfixed} 
G_{\lambda,\kappa}(l_{in},l_{out};1)=e^{-\lambda\,a^2\,(l_{in}+l_{out})}\sum_{\genus=0}^{[N/2]}
\binom{N}{l_{in}}\binom{N}{2\genus}\frac{(2\genus)!}{\genus !
(\genus+1)!}e^{-2\kappa\genus}.
\end{equation}
From this one can obtain the propagator $G_{\lambda,\kappa}(l_{in},l_{out};t)$ for arbitrary time $t$ by iterating \eqref{eq:toptrans:sumfixed} $t$ times according to the composition law
 \begin{eqnarray}
\label{eq:topotrans:composition}
G_{\lambda,\kappa}(l_{in},l_{out},t_1+t_2)&=& \sum_{l}G_{\lambda,\kappa}(l_{in},l,t_1)
G_{\lambda,\kappa}(l,l_{out},t_2),\\
G_{\lambda,\kappa}(l_{in},l_{out},t+1)&=& \sum_{l}G_{\lambda,\kappa}(l_{in},l,1)
G_{\lambda,\kappa}(l,l_{out},t).
\end{eqnarray}
To give a complete solution of the discrete problem we still have to explicitly perform the sum over all genus in \eqref{eq:toptrans:sumfixed}. We will obtain this result in  Laplace transformed ``momentum'' space, by introducing generating functions for the one-step propagator
\begin{equation}
\label{eq:topotrans:gen} 
G(x,y;g,h;1)=\sum_{l_{in},l_{out}=0}^\infty
G_{\lambda,\kappa}(l_{in},l_{out},1) x^{l_{in}} y^{l_{out}},
\end{equation}
where we have defined $g\equ e^{-\lambda }$, as in the pure model, and $h\equ e^{-\kappa}$. Upon inserting \eqref{eq:toptrans:sumfixed} into \eqref{eq:topotrans:gen} and evaluating the summations over the $l$'s, one obtains the generating function of the one-step propagator
\begin{equation}
\label{eq:topotrans:Glaplace}
 G(x,y,g,h,1)=    \frac{1}{1 - g\,\left( x + y \right) }
 \frac{2}{1 + {\sqrt{1 - 4 u^2}}},
\end{equation}
with
\begin{equation} \label{z}
  u = \frac{h}{1 - \frac{1}{g\,\left( x + y \right) }}.
\end{equation}
Note that in order to arrive at this expression,
we have explicitly performed the sum over all (not-so-bad) topologies! 
The fact that this infinite sum converges for appropriate values of
the bare couplings has to do with the restriction to the ``not-so-bad'' topology changes according to the imposed causality constraints, as motivated in the previous section.

In \eqref{eq:topotrans:Glaplace} one recognizes the generating function 
$\Cat(u^2)$ for the Catalan numbers,
\begin{equation}\label{eq:topcont:catalan}
 \Cat(u^2)=\frac{2}{1 + {\sqrt{1 - 4 u^2}}}=\frac{1-\sqrt{1-4 u^2}}{2u^2}.
\end{equation}
For $h\equ 0$ one has $\Cat(u^2)\equ 1$ and expression 
\eqref{eq:topotrans:Glaplace} reduces to the generating function of the
one-step propagator with open spatial boundary conditions of the pure CDT model (cf. Appendix \ref{App:boundary}),
\begin{equation}
G(x,y;g,h=0;1)=\frac{1}{1 - g x -g y }. \label{eq:topotrans:without}
\end{equation}

\subsection{Continuum and double scaling limit}\label{sec:topocont}

Taking the continuum limit in the case of pure CDT is
fairly straightforward, as we have seen in Section \ref{sec:cont}. The joint region of convergence of the one-step propagator \eqref{eq:topotrans:without} is given by
\begin{equation}
|x|<1,\quad |y|<1,\quad |g|<\frac{1}{2}.
\end{equation}
The continuum limit was then obtained by simultaneously fine-tuning to the critical values with use of the following canonical scaling relations
\begin{eqnarray}
g & = & \frac{1}{2}(1-a^2\,\Lambda) + \mathcal{O}(a^{3}),\label{eq:topcont:scalinga}\\
x & = & 1-a\,X + \mathcal{O}(a^{2}),\label{eq:topcont:scaling2}\\
y & = & 1-a\,Y + \mathcal{O}(a^{2}),\label{eq:topcont:scaling3}
\end{eqnarray}
where $X$, $Y$ and $\Lambda$ denoted the renormalized couplings. The difficulty which arises when taking the continuum limit in the case with topology changes is to find a suitable  \textit{double scaling limit} for both the gravitational and cosmological coupling, which leads to a \textit{physically sensible continuum theory}.  By
this we mean that the one-step propagator should
yield the Dirac delta-function to lowest order in $a$, and that the
Hamiltonian should be bounded below and not
depend on higher-order terms of the scaling relations for all couplings,
in a way that would introduce a non-trivial dependence on new couplings in the continuum theory.
Difficulties in finding such a double scaling limit arise due to the fact that Newton's constant is dimensionless in two
dimensions, whence there is no preferred canonical scaling for $h$. 
One can make the multiplicative ansatz\footnote{Here the factor
$\frac{1}{\sqrt{2}}$ is chosen to give a proper parametrization of
the number of holes in terms of Newton's constant (see Section
\ref{sec:topobserv}).}
\begin{equation}
\label{scalingh} h =\frac{1}{\sqrt{2}} h_{ren} (a d)^\beta,
\end{equation}
where $h_{ren}$ depends on the renormalized Newton's constant
$G_N$ according to 
\begin{equation}
\label{hrendef}
h_{ren}=e^{- 2\pi / G_N}.
\end{equation}
In order to compensate the powers of the cut-off $a$ in (\ref{scalingh}),
$d$ must have dimensions of inverse length.
The most natural ansatz in terms of the dimensionful quantities 
available is
\begin{equation} 
\label{d}
d = (\sqrt{\Lambda}^{\alpha} (X+Y)^{1-\alpha}),
\end{equation}
where the constants $\alpha$ and $\beta$ have to be chosen such that one obtains a physical sensible continuum theory according to the above considerations.

To calculate the effective quantum Hamiltonian one can follow a similar procedure as the one used in the case of the pure CDT model in Section \ref{sec:cont}. Thereby, we start off with the one-step time evolution of the discrete wave function, recall \eqref{eq:cont:timeevol},
\begin{equation}
\label{eq:topcont:timeevol}
\psi(x,t+1)= \oint \frac{dz}{2 \pi i z} G(x,z^{-1};g,h;1)\,\psi(z,t).
\end{equation}
Upon inserting the scaling relations \eqref{eq:topcont:scalinga}-\eqref{eq:topcont:scaling3} and 
$t\equ \frac{T}{a}$ into this equation and using
\begin{equation}
\psi(X,T+a)=e^{-a\,\hat{H}(X,\partial_X)}\psi(X,T),
\end{equation}
one can expand both sides to first order in a, yielding
\begin{equation}
\label{eq:toptcont:scaledtransfernog}
\left(1-a \hat{H}+\mathcal{O}(a^2)\right)\psi(X) = \int^{i\infty}_{-i\infty} 
\frac{dZ}{2\pi i} \left\{\left(\frac{1}{Z-X} + a\,\frac{2\Lambda - X Z}{(Z-X)^2}\right)
\Cat(u^2) \right\} \psi(Z),
\end{equation}
with $\psi(X)\equiv\psi(x\equ 1- a X)$.
For convenience, we treated separately the
first factor in the one-step propagator \eqref{eq:topotrans:Glaplace}, which is nothing but the expansion of the
one-step propagator without topology changes, and
the second factor, the Catalan generating function \eqref{eq:topcont:catalan}, which contains all information on the new couplings. Note that the first term on the right-hand side of \eqref{eq:toptcont:scaledtransfernog}, $\frac{1}{Z-X}$, 
is the Laplace-transformed delta-function. This gives us sensible information of the values of $\alpha$ and $\beta$ in the expansion of the Catalan generating function. Inserting the scaling relations \eqref{eq:topcont:scalinga}-\eqref{scalingh} into $\Cat(u^2)$ and expanding, one obtains
\begin{equation} \label{eq:topcant:catexp}
\Cat(u^2) = 1 + \frac{2\,d ^{2\beta }\,h_{ren}^2}{\,{\left( Z-X \right) }^2}\,
a^{2\beta-2} + \mathcal{O} (a^{2\beta-1}).
\end{equation}
Thus, in order to preserve the delta-function to lowest order in \eqref{eq:toptcont:scaledtransfernog} and to have a
non-vanishing contribution to the Hamiltonian one is naturally led to
$\beta \equ 3/2$. For suitable choices of $\alpha$ it is also possible to obtain 
the delta-function by setting $\beta\equ 1$, but
the resulting Hamiltonians turn out to be unphysical or at least do not
have an interpretation as gravitational models with wormholes, 
as we will discuss in Appendix \ref{app:scaling}.\footnote{One might also consider scalings of the form 
$ h \rightarrow  c_1 h_{ren} (a d)+c_2 h_{ren} (a d)^{3/2}$, but
they can be discarded by arguments similar to those of Appendix \ref{app:scaling}.
}

Considering $\beta \equ 3/2$ and inserting \eqref{eq:topcant:catexp} into \eqref{eq:toptcont:scaledtransfernog}, the right hand side of this equation becomes
\begin{equation} 
\label{scaledtransfernognew}
 \int^{i\infty}_{-i\infty} \frac{dZ}{2\pi i} \left\{\frac{1}{Z-X} + a\,\left( \frac{2\Lambda - X Z}
{(X-Z)^2}-\frac{2\,\sqrt{\Lambda}^{3 \alpha} \,h_{ren}^2}
{\,{\left( X-Z \right) }^{3 \alpha}} \right)\right\} \psi(Z).
\end{equation}
Observe that to have a non-trivial dependence on the new couplings in the continuum theory, 
one should not allow for values $\alpha\klgl 0$. Hence, performing the integration for $\alpha\gro 0$ and discarding possible fractional poles in \eqref{scaledtransfernognew}, the effective quantum Hamiltonian in ``momentum'' space reads
\begin{equation}
\label{HamiltonianX}
\hat{H}(X,\frac{\partial}{\partial X})=X^2 \frac{\partial}{\partial X} + 
X -2 \Lambda \frac{\partial}{\partial X}+
2\Lambda^{\frac{3 \alpha}{2}}h_{ren}^2
\frac{(-1)^{3\alpha}}{ \Gamma(3\alpha)} \frac{\partial^{3\alpha-1}}{\partial X^{3\alpha-1}},\quad \alpha=\frac{1}{3},\frac{2}{3},1,...\, .
\end{equation}
For all possible values of $\alpha$, these Hamiltonians do not depend on higher order terms in the scaling relations \eqref{eq:topcont:scalinga}-\eqref{scalingh}. One might check this, as explained in Section \ref{sec:cont}, by explicitly introducing higher order terms in the scaling relations and observing that the resulting Hamiltonians are independent of them. 

One can now perform the inverse Laplace transformation on the wave function, $\mbox{$\psi(L)\equ\int^{i\infty}_{-i\infty} \frac{dX}{2\pi i} e^{X\,L}\psi(X)$}$, to obtain the effective quantum Hamiltonian in length space,
\begin{equation}
\label{Hamiltonian}
\hat{H}(L,\dL)=-L\ddL-\dL+2\,\Lambda\,L- \frac{2\,\Lambda^\frac{m+1}{2}h_{ren}^2 }
{ \Gamma(m+1)} L^m,\quad m=0,1,2,...\, ,
\end{equation}
where we defined $m\equ 3\alpha\mi 1$. Since $\hat H$ is unbounded below for $m \grgl 2$, we are left
with $m\equ 0$ and $m\equ 1$ as possible choices for the scaling. 
However, setting $m\equ 0$ merely has the effect of adding a constant term 
to the Hamiltonian, leading to a trivial phase factor for the wave function. Hence, we conclude that the only continuum theory with non-trivial dependence on the new couplings corresponds to 
$m\equ 1$, i.e. to the following double scaling limit
\begin{equation}
\label{hm1}
h^2=\frac{1}{2} h_{ren}^2\,\Lambda\,(X+Y)\,a^3,
\end{equation}
with the Hamiltonian given by
\begin{equation}
\label{Hamiltonian1}
\hat{H}(L,\dL)=-L\ddL-\dL+  \left(1- h_{ren}^2 \right)\,2\,\Lambda\,L.
\end{equation}
Note that for all values $G_N\geq 0$ of the renormalized
Newton's constant \eqref{scalingh} the Hamiltonian is bounded from
below and therefore well defined. Further, the Hamilton operator is selfadjoint on the Hilbert space $\mathcal{H}=\mathcal{L}^2(\mathbb{R}_+,dL)$. Note that \eqref{Hamiltonian1} corresponds to the propagator with open boundary conditions. It is not difficult to recover the  Hamilton operator for the case of circular boundary conditions
\begin{equation}
\label{Hamiltonian2}
\hat{H}(L,\dL)=-L\ddL-2 \dL+  \left(1- h_{ren}^2 \right)\,2\,\Lambda\,L.\quad (\text{topology }S^1\!\times\![0,1]),
\end{equation} 
which is selfadjoint on the Hilbert space $\mathcal{H}=\mathcal{L}^2(\mathbb{R}_+,L dL)$.
The spectrum of this Hamiltonian is discrete with equidistant eigenvalues,
\begin{equation}
\label{eq:topcont:Energy}
E_n=2\sqrt{2 \Lambda(1-h_{ren}^2)}(n+1)\quad n=0,1,2,..
\end{equation}
The corresponding eigenfunctions are given by
\begin{equation}
\label{eq:topcont:eigenfunc}
\psi_n(L)=\mathcal{A}_n e^{-\sqrt{2\Lambda(1-h_{ren}^2)}L}\, {}_1F_{1}(-n;2;2\sqrt{2\Lambda(1-h_{ren}^2)}L),\quad d\mu(L)=LdL,
\end{equation}
where ${}_1F_{1}(-n,\mu+1;z)$ are the confluent hypergeometric functions, defined in \eqref{eq:cont:hypergeom}. 
As in the case of pure CDT, the eigenfunctions $\psi_n(L)$ form an orthonormal basis of the Hilbert space $\mathcal{H}$, where the normalization factor is given by
\begin{equation}
\mathcal{A}_n=2\sqrt{2\Lambda(1-h_{ren}^2)(n+1)}.
\end{equation}

Having obtained the spectrum of the Hamiltonian one can easily calculate the (Euclidean) partition function
\begin{equation}
\label{eq:topcont:partitionfuncres}
\mathcal{Z}_T(\Lambda,G_N)=\sum_{n=0}^\infty e^{-TE_n}=\frac{1}{e^{2\sqrt{2\Lambda (1-h_{ren}^2)} T} -1},\quad h_{ren}=e^{- 2\pi / G_N}.
\end{equation}
Further the time propagator or loop-loop correlator can be obtained as (cf. Appendix \ref{App:Calogero})
\begin{eqnarray}
    G_{\Lambda,G_N}(L_1,L_2;T)&=&\sum_{n=0}^{\infty}e^{-TE_n}\psi_n^*(L_2)\psi_n(L_1)\\
    &=& \frac{\omega} {\sqrt{L_1L_2}}
	\frac{e^{-\omega (L_1+L_2)\coth(\omega T)}}{\sinh(\omega T)}
	I_1 \left(\frac{2\omega\sqrt{ L_1 L_2}}{\sinh(\omega T)}\right),
\end{eqnarray}
where we have used the short hand notation $\omega\equ\sqrt{2\Lambda(1-
h_{ren}^2)}$. As expected, for $h_{ren}\!\rightarrow\! 0$ the results
reduce to those of the pure two-dimensional CDT model, as obtained in Section \ref{sec:cont}.

\subsection{Physical observables}\label{sec:topobserv}

In the following we want to analyze the new model of CDT with topology changes in terms of its physical observables. As in the pure model, there are certain geometric observables such as the expectation values of the spatial length and higher moments, $\mbox{$\expec{L^m}_n\ident\bra{n}L^m\ket{n}$}$, which can be calculated using the expressions for the wave functions \eqref{eq:topcont:eigenfunc}, as obtained in the previous section. The first results read
\begin{equation}
\label{eq:topobs:length}
\expec{L}_n=\frac{n+1}{\sqrt{2\Lambda(1-
h_{ren}^2)}},\quad\expec{L^2}_n=\frac{3}{4}\frac{(n+1)^2}{\Lambda(1-
h_{ren}^2)},
\end{equation}
where a general expression for higher moments is presented in Appendix \ref{App:Calogero}. Generally, the moments scale as $\expec{L^m}_n=\Lambda^{-\frac{m}{2}}$ as in the pure CDT model. From \eqref{eq:topobs:length} one can then also calculate the variance $\expec{\Delta L}_n$ of the spatial length,
\begin{equation}
\expec{\Delta L}_n=\sqrt{\expec{L^2}_n-\expec{L}_n^2}=\frac{1}{2}\,\frac{n+1}{\sqrt{\Lambda(1-
h_{ren}^2)}}.
\end{equation}
One can still interpret these observables in terms of a fluctuating universe, as done in the pure CDT model, where the quantum fluctuations are now governed by the ``effective'' cosmological constant $\Lambda_{\mathrm{eff}}\equ\Lambda (1\mi
h_{ren}^2)$. A detailed discussion on this point in terms of some new ``topological'' observables will be given in the next section.

Another geometrical observable is the Hausdorff dimension $d_H$, which can be calculated from the ensemble average over all geometries with arbitrary genus. Since the partition function scales as 
\begin{equation}
\mathcal{Z}_T(\Lambda,G_N) \sim e^{2\sqrt{2\Lambda (1-h_{ren}^2)} T}\quad \text{as } T\rightarrow\infty,
\end{equation}
a similar calculation, like the one performed in Section \ref{sec:obs}, reveals that the Hausdorff dimension is given by
\begin{equation}
d_H=2.
\end{equation} 
This is the same result as obtained in the case of pure two-dimensional CDT.

\begin{figure}
\begin{center}
\includegraphics[width=4in]{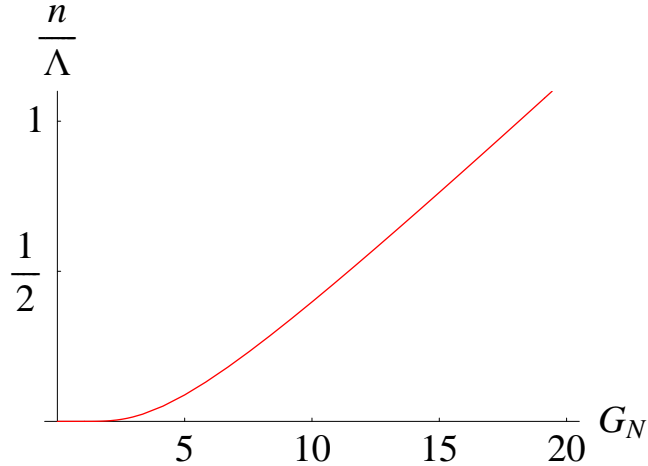}
\caption{The density of wormholes $n$ in units of $\Lambda$ as a
function of Newton's constant $G_N$.}\label{fig:densityholes}
\end{center}
\end{figure}

In addition to these well-known geometric observables, 
the system possesses a new type of ``topological" observable
which involves the number of wormholes $N_\genus$, as already anticipated in Section \ref{sec:implemholes}, where we mentioned that the presence of 
wormholes in the quantum geometry and
their density can be determined from light scattering.
An interesting quantity to
calculate is the average number of wormholes in a piece of spacetime
of duration $T$, with initial and final spatial boundaries identified. 
Because of the simple dependence of the action on the genus
this is easily computed by taking the derivative of the
partition function $\mathcal{Z}_T(\Lambda,G_N)$ with respect to the corresponding
coupling, namely,
\begin{equation}
\label{Nholedefpartfunc}
    \expec{N_\genus} = \frac{1}{\mathcal{Z}_T(\Lambda,G_N)}\frac{h_{ren}}{2}\frac{\partial\, \mathcal{Z}_T(\Lambda,G_N)}{\partial h_{ren}}.
\end{equation}
Upon inserting \eqref{eq:topcont:partitionfuncres} this yields
\begin{equation}\label{Nholeresultpartfunc}
    \expec{N_\genus} =  T \, h_{ren}^2 \Lambda\, \frac{\coth \left(\sqrt{2\Lambda
    (1-h_{ren}^2)}\,T\right)-1}{ \sqrt{2\Lambda (1-h_{ren}^2)}}.
\end{equation}
In an analogous manner we can also calculate the average spacetime volume
\begin{equation}\label{averagevoldef}
    \expec{V} = -\frac{1}{\mathcal{Z}_T(\Lambda,G_N)}\frac{\partial\, \mathcal{Z}_T(\Lambda,G_N)}{\partial \Lambda},
\end{equation}
leading to
\begin{equation}\label{averagevolresult}
    \expec{V} = T \, \sqrt{\frac{(1-h_{ren}^2)}{2 \Lambda}}\left(
\coth\left(\sqrt{2\Lambda(1-h_{ren}^2)}\,T\right)-1\right).
\end{equation}
Dividing \eqref{Nholeresultpartfunc} by \eqref{averagevolresult} we find 
that the spacetime density of wormholes $n$ is finite,
\begin{equation}
\label{densityfinalresult}
n=\frac{\expec{N_\genus}}{\expec{V}}= \frac{h_{ren}^2}{1-h_{ren}^2} \,\Lambda.
\end{equation}
The density of holes in terms of the renormalized Newton's constant is given by
\begin{equation}
\label{densityonG}
n=\frac{1}{e^\frac{4\pi}{G_N}-1} \,\Lambda.
\end{equation}
The behaviour of $n$ in terms of the renormalized Newton's constant is shown 
in Figure \ref{fig:densityholes}. The density of holes vanishes as $G_N\!\rightarrow\! 0$ and 
the model reduces to the case without topology change.

\subsection{Taming the cosmological constant}\label{sec:topotaming}

We have already seen in the last section that fluctuations in the geometry gave rise to the notion of the ``effective'' cosmological constant $\Lambda_{\mathrm{eff}}\equ\Lambda (1\mi
h_{ren}^2)$. In the following we want to interpret this effect in terms of the physical quantities, namely, the cosmological scale $\Lambda$ and the density of wormholes in units of $\Lambda$, i.e. $\eta=\frac{n}{\Lambda}$. These can be seen as the two scales of the model, in contrast to the pure CDT model which just has a single scale. As in the latter, 
the cosmological constant, as the only dimensionful quantity, defines the global length scale of the two-dimensional ``universe" through $\expec{L}\sim\frac{1}{\sqrt{\Lambda}}$. 
The new scale in the model with topology change is the relative scale $\eta$ 
between the cosmological and topological fluctuations, which is parametrized
by Newton's constant $G_N$. Both together govern the effective fluctuations in the quantum geometry.  

\begin{figure}
\begin{center}
\includegraphics[width=4in]{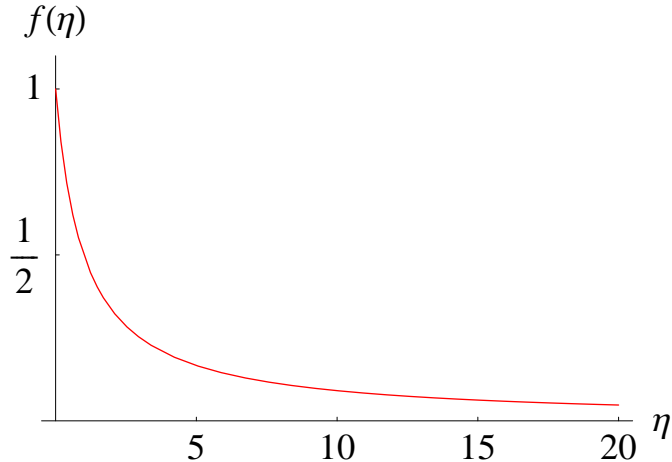}
\caption{The coefficient of the effective potential, $f(\eta)=1/(1+\eta)$,
as function of the density of holes in units of $\Lambda$, 
$\eta=\frac{n}{\Lambda}$.}\label{fig:effpot}
\end{center}
\end{figure}

One can nicely observe this when rewriting and interpreting the Hamiltonian \eqref{Hamiltonian1} in 
terms of the new physical 
quantities, $\Lambda$ and $\eta$, resulting in
\begin{equation}
\label{hamiltonianeta}
\hat{H} (L,\dL)=-L\ddL-\dL+  \frac{1}{1+\eta} \,2\,\Lambda\,L.
\end{equation}
One sees explicitly that the topology fluctuations affect the dynamics since the 
effective potential depends on $\eta$, as illustrated by Figure \ref{fig:effpot}. From the coefficient of the effective potential, $f(\eta)=1/(1-\eta)$, one also observes that the presence of wormholes in our model leads to a decrease of the ``effective" cosmological constant $\Lambda_{\mathrm{eff}}\equ f(\eta)\Lambda$. 

This connects nicely to former attempts to 
devise a mechanism, the so-called \textit{Coleman's mechanism}, to explain the smallness of the cosmological constant 
in the Euclidean path integral formulation of {\it four}-dimensional quantum gravity 
in the continuum with the presence of infinitesimal wormholes \cite{Coleman:1988tj,Klebanov:1988eh}. The wormholes considered there resemble those of our toy model 
in that both are non-local identifications of the spacetime geometry of 
infinitesimal size. The counting of our wormholes is of course different since 
we are working in a genuinely Lorentzian setup where certain causality 
conditions have to be fulfilled. This enables us to do the sum over
topologies explicitly.

Further, in Coleman's mechanism for driving the cosmological constant $\Lambda$ to zero, an additional sum over different 
baby universes is performed in the path integral, which leads to a 
distribution of the cosmological constant that is peaked near zero. 
We do not consider such an additional sum over baby universes, 
but instead have an explicit expression for the effective potential which 
shows that an increase in the number of wormholes is accompanied by a 
decrease of the ``effective" cosmological constant. \\

To conclude, we have seen that two-dimensional 
Lorentzian quantum gravity including a sum over topologies leads to a well-defined unitary continuum quantum theory, different to the one of the pure CDT model. The presence of causality constraints imposed on the
path-integral histories -- physically motivated in
Section \ref{sec:implemholes} --
enabled us to derive a new class of continuum theories by taking an
unambiguously defined double-scaling limit of a statistical
model of simplicially regularized space-times. These causality constraints were crucial to resolve the problem of super-exponential growth in the number of configurations, as present in the model of two-dimensional Euclidean quantum gravity with topology changes.  
The resulting model of two-dimensional CDT with topology changes has, besides the well-known geometrical observables, new ``topological'' observables, such as the finite space-time density of wormholes. Further, we observed that the presence of wormholes in our model leads to a decrease of the ``effective" cosmological constant, which nicely connected to former attempts of driving the cosmological constant $\Lambda$ to zero in the formulation of {\it four}-dimensional quantum gravity 
in the continuum through the presence of infinitesimal wormholes, as described by Coleman's mechanism.

\section{Universality: Its physical and mathematical implications} \label{sec:univers}

In this section we discuss generalizations and implications of the ``pure'' two-dimensional CDT model, as described in Section \ref{sec:CDT}. 
In doing this, we use the techniques of the formalism developed in Section \ref{sec:CDT}.
The section is structured as follows: In Section \ref{sec:higher} we introduce a higher curvature term in the pure model and show that the resulting theory is equivalent to a two-dimensional Lorentzian quantum gravity model whose discrete space-times consist of squares and triangles. Further we show that it belongs to the universality class of the pure two-dimensional CDT model. In Section \ref{sec:limited} a model is introduced which only allows for minimal curvature weights per space-time strip, where we see that the resulting model also shares the same universality class with the pure two-dimensional CDT model. 
This makes it clear that two-dimensional CDT define a rather broad universality class and is thus a genuine continuum theory independent of the details of the regularization.
In Section \ref{sec:dimer} we discuss bijections between CDTs, heaps of dimers and Dyck paths. Such connections can be a useful mathematical tool to reduce the two-dimensional counting problem to an effective one-dimensional problem.

\subsection{2D Lorentzian quantum gravity with a higher curvature term}\label{sec:higher}

Following \cite{DiFrancesco:1999em} we introduce a higher curvature term in the pure model of 2D Lorentzian quantum gravity. We will see that the resulting theory preserves universality of the pure model. Further, the higher curvature term can be interpreted as being equivalent to inserting squares in the triangulation of the pure model.

In the pure two-dimensional CDT model we attached a ``cosmological'' weight $g\equ e^{-\lambda}$ to every triangle face resulting to the factor $g^{N(T)}$ in the Boltzmann weight of each triangulation $T$, where $N(T)\equ\lin\pl\lout$ is the number of triangles in the triangulation. Since the usual curvature term in the Einstein-Hilbert action is trivial in 2d for fixed topology, there was no curvature weight in the propagator. 

We now want to generalize the pure model by explicitly introducing a higher curvature weight into the propagator which suppresses or enhances local curvature. To still be able to use the powerful techniques of the transfer matrix formalism, we require that this higher curvature weight is defined locally within one strip of height $\Delta t\equ1$. A convenient way to define such a higher curvature weight is by assigning a weight $\Theta$ to each pair of adjacent triangles pointing both up or pointing both down \cite{DiFrancesco:1999em,Kazakov:1995gm}. The resulting effect of this higher curvature term in the eigenvalues of the transfer matrix can most easily be calculated using the dual graphical representation of the triangulation. 
\begin{figure}
\begin{center}
\includegraphics[width=5in]{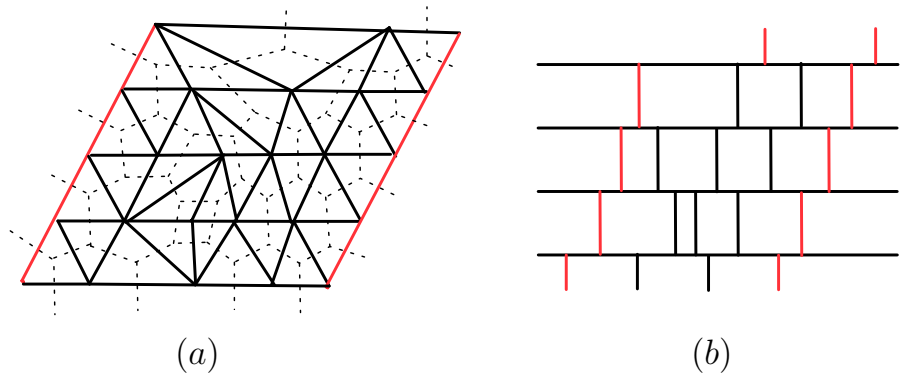}
\caption{(a) A typical triangulation with staircase boundary conditions. The dashed lines show the relation to the corresponding dual triangulation which is drawn in (b).}\label{fig:dual}
\end{center}
\end{figure}
For convenience we consider a triangulation with ``staircase'' boundary conditions\footnote{The resulting theory for ``staircase'' boundary conditions is closely related to the one with periodic boundary conditions as shown in Appendix \ref{App:boundary}.}, as illustrated in Figure \ref{fig:dual} (a). In the dual representation (b) a space-time strip of height $\Delta t\equ 1$ translates into a sequence of half-edges attached to the dual constant time line, where the half-edges above (below) this line correspond to the up-pointing (down-pointing) triangles in the triangulation (a). In terms of the dual triangulation we assign a cosmological weight $g$ to every vertex (which is in fact the dual of the triangle face) and in addition each pair of neighboring up-pointing half-edges or down-pointing half-edges receives a factor $\Theta$,
\begin{equation}
\raisebox{-25pt}{\includegraphics[height=65pt]{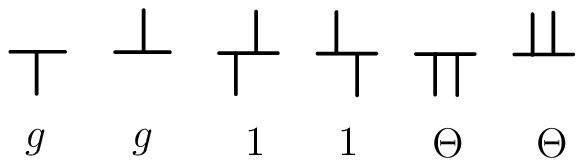}}.
\end{equation}

Using these weights we can calculate the eigenvalues of the transfer matrix $G_{g,\Theta}(l_{in},l_{out};t\equ 1)\equ\bra{l_{out}}\Trans\ket{l_{in}}$ by summing over possible configurations of a dual time line with $\lin$ half-edges below and $\lout$ half-edges above. In spirit of the higher curvature weight this is done by first summing over the number $k\!\geqslant\! 0$ of blocks of $n_r\!\geqslant\! 1$ neighboring lower half-edges and $m_r\!\geqslant\! 1$ neighboring upper half edges, with $r=1,...k$, and summing over all partitions of lower half-edges, $\lin=\sum_{r}n_r$, and upper half-edges , $\lout=\sum_{r}m_r$, yielding  
\begin{equation}
G_{g,\Theta}(l_{in},l_{out};t= 1)=(g\,\Theta)^{\lin+\lout}\sum_{k\geqslant 1}\frac{1}{\Theta^{2k}}\binom{\lin-1}{k-1}\binom{\lout-1}{k-1}.
\end{equation}
Introducing the generating function on the propagator $G_{g,\Theta}(l_{in},l_{out};t\equ 1)$ one obtains
\begin{eqnarray}
G(x,y;g,\Theta;1) & = & \sum_{l_{in},l_{out}}G_{g,\Theta}(l_{in},l_{out};t\equ 1) \nonumber\\
 & = & \frac{g^2 x y}{1-g\,\Theta(x+y)-g^2(1-\Theta^2)xy}\label{eq:higher:gen}. 
\end{eqnarray}
As expected, for $\Theta\equ1$ one recovers the one-step propagator of the pure model  \eqref{eq:appbound:staircase}, whereas for $\Theta\equ 0$ curvature is fully suppressed and one gets the one-step propagator for a flat triangulation.

It is very interesting to see that the discrete one step-propagator \eqref{eq:higher:gen} can be reinterpreted as the one-step propagator of a CDT model whose discrete space-times consists of triangles and squares.

\begin{figure}
\begin{center}
\includegraphics[width=5in]{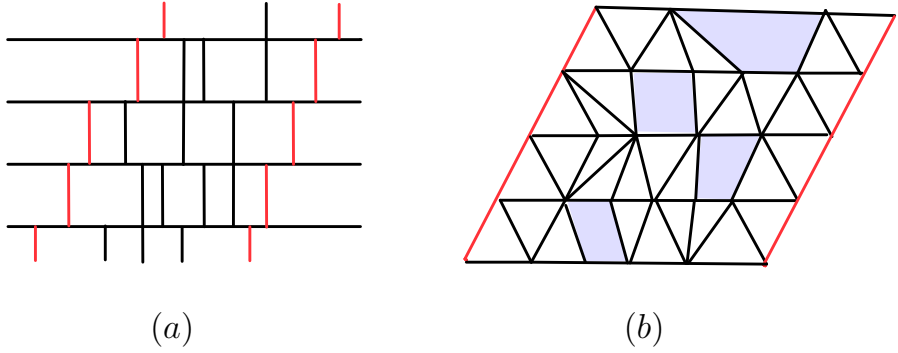}
\caption{Reinterpretation of the resulting model with higher curvature terms. We assign a weight $g\,\Theta$ to isolated half-edges, $\bot$ and $\top$, and a weight $g^2(1-\Theta^2)$ to crossings $+$. A typical triangulation corresponding to these weights is drawn in (a). The dual triangulation (b) consists of up- and down-pointing triangles and squares.}\label{fig:dualsquare}
\end{center}
\end{figure}

Expanding the denominator of \eqref{eq:higher:gen}, one only gets sequences of the terms $g\Theta x$, $g\Theta y$ and $g^2(1-\Theta^2)$. We can now assign a factor $g\Theta x$ ($g\Theta y$) to a lower (upper) half-edge which both have a weight $g\Theta$. The factor $g^2(1-\Theta^2)$ then belongs to a pair of upper and lower half-edge which can be merged to a crossing. The resulting new weights read,
\begin{equation}
\raisebox{-25pt}{\includegraphics[height=65pt]{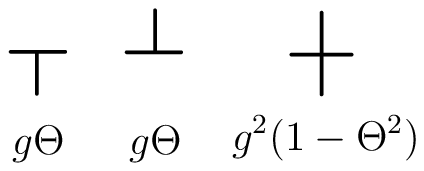}}.
\end{equation}
The corresponding space-time is shown in Figure \ref{fig:dualsquare} (a). The dual of this space-time is made up of triangles and squares, as illustrated in (b). In the following we want to look at the continuum theory of such a two-dimensional Lorentzian quantum gravity model whose space-times are made up of triangles and squares in the discrete setting.

Performing the continuum limit of \eqref{eq:higher:gen} is straightforward. The joint region of convergence of \eqref{eq:higher:gen} is given by
\begin{equation}
|x|<1,\quad |y|<1,\quad |g|<\frac{1}{1+\Theta}.
\end{equation}
One can fine-tune to the critical point $(x_c,y_c,g_c)=(1,1,\frac{1}{1+\Theta})$ by use of the following scaling relations
 \begin{eqnarray}
g & = & \frac{1}{1+\Theta}(1-a^2\,\Lambda) + \mathcal{O}(a^{3}),\label{eq:higher:1}\\
x & = & 1-a\,X + \mathcal{O}(a^{2}),\label{eq:higher:2}\\
y & = & 1-a\,Y + \mathcal{O}(a^{2}).\label{eq:higher:3}
\end{eqnarray}
Upon inserting \eqref{eq:higher:1}-\eqref{eq:higher:3} into \eqref{eq:higher:gen} and following the formalism developed in Section \ref{sec:CDT} one obtains the effective quantum Hamiltonian of the resulting continuum theory,
\begin{equation}
\hat{H}(L,\dL)=-\Theta L\ddL-\Theta \dL +2\Lambda L,
\end{equation}
which is self-adjoint on the Hilbert space $\mathcal{H}\equ \mathcal{L}^2(\mathbb{R}_+,dL)$. Further, using \eqref{calogero20}, one can read off the finite time propagator
\begin{eqnarray}
	G_{\Lambda,\Theta}(L_1,L_2,T)=\sqrt{\frac{\Lambda}{\Theta}}
	\frac{e^{-\sqrt{\frac{\Lambda}{\Theta}}(L_1+L_2)\coth(\sqrt{\Theta\Lambda}T)}}{\sinh(\sqrt{\Theta\Lambda}T)}
	I_0\left(\frac{2}{\sqrt{\Theta}}\frac{\sqrt{\Lambda L_1 L_2}}{\sinh(\sqrt{\Theta\Lambda}T)}\right).
\end{eqnarray}
One sees that the resulting higher curvature model is equivalent to the pure CDT model (with staircase boundary conditions) up to a global rescaling of the coordinates, $\Lambda\!\mapsto\! \frac{\Lambda}{\Theta}$, $L_1\!\mapsto\! \Theta L_1$ and $L_2\!\mapsto\! \Theta L_2$. 
Hence, the new discretization of the two-dimensional Lorentzian quantum gravity model whose space-times are made up out of triangles and squares belongs to the same universality class as the pure CDT model which is discretized with triangles only.

In \cite{DiFrancesco:1999em} it is shown that one can generalize this treatment to discrete space-times which incorporate 
p-polygons, where the resulting continuum theory still belongs to the same universality class as the pure two-dimensional CDT model.

\subsection{2D Lorentzian quantum gravity with minimal curvature weights}\label{sec:limited}

In this section we want to discuss another generalization of the pure two-dimensional CDT model, where one only allows for minimal (positive or negative) curvature weights per space-time strip of height $\Delta t\equ 1$. By this we mean that the number of neighboring triangles pointing in the same direction is not allowed to exceed two.

In terms of the discrete setting the construction is very simple: to calculate the eigenvalues of the transfer matrix, $G_{\Lambda}(l_{in},l_{out};t\equ 1)\equ\bra{l_{out}}\Trans\ket{l_{in}}$, one sums over all possible triangulations of a space-time strip of height $\Delta t\equ 1$ with $\lin$ up-pointing triangles and $\lout$ down-pointing triangles, where in addition one restricts the maximal number of neighboring triangles pointing in the same direction to $N_{max}\equ 2$. In the continuum limit this means that we only consider an infinitesimal amount of curvature per infinitesimal space-time strip. However, we expect that in a small, but finite region of space-time we can still generate arbitrarily large values of curvature. More precisely, we expect that the resulting continuum theory will be equivalent to the one obtained in Section \ref{sec:higher}, with a fixed curvature suppressing factor $\Theta\!\in\! (0,1)$.

To solve the discrete problem we introduce the generating function for the one-step propagator $G_{\Lambda}(l_{in},l_{out};t\equ 1)$. Hereby one can take advantage of graphical methods, where one assigns a factor of $gx$ to every up pointing triangle ``$\bigtriangleup$'' and a factor of $gy$ to every down pointing triangle ``$\bigtriangledown$''. The one-step propagator for the case of open boundary conditions then reads
\begin{eqnarray}
G(x,y;g;1)&=&\sum_{k=0}^\infty \left(
\raisebox{-10pt}{\includegraphics[height=25pt]{route1}}+
\raisebox{-10pt}{\includegraphics[height=25pt]{route2}}+
\raisebox{-10pt}{\includegraphics[height=25pt]{route3}}+
\raisebox{-10pt}{\includegraphics[height=25pt]{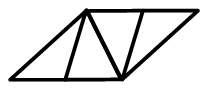}}
 \right)^k\nonumber\\
 &=&\frac{1}{1-g^2 x y (1+gx)(1+gy)}.\label{eq:limited:gen}
\end{eqnarray}
The joint region of convergence of \eqref{eq:limited:gen} is given by
 \begin{equation}
|x|<1,\quad |y|<1,\quad |g|<\frac{1}{2}(\sqrt{5}-1).
\end{equation}

To perform the continuum limit we fine-tune to the critical point $\mbox{$(x_c,y_c,g_c)\equ(1,1,\frac{1}{2}(\sqrt{5}-1))$}$ by use of the following scaling relations
\begin{eqnarray}
g & = & \frac{1}{2}(\sqrt{5}-1)(1-a^2\,\Lambda) + \mathcal{O}(a^{3}),\label{eq:limited:1}\\
x & = & 1-a\,X + \mathcal{O}(a^{2}),\label{eq:limited:2}\\
y & = & 1-a\,Y + \mathcal{O}(a^{2}).\label{eq:limited:3}
\end{eqnarray}
Inserting the scaling relations \eqref{eq:limited:1}-\eqref{eq:limited:3} into the one-step propagator \eqref{eq:limited:gen} and following the standard procedure one obtains the effective quantum Hamiltonian of the resulting continuum theory,
\begin{equation}
\hat{H}(L,\dL)=-\Theta L\ddL-\Theta \dL +2\Lambda L,\quad \text{with }\Theta=\frac{\sqrt{5}-1}{\sqrt{5}+5}\approx 0.17,
\end{equation}
which is self-adjoint on the Hilbert space $\mathcal{H}\equ \mathcal{L}^2(\mathbb{R}_+,dL)$. One sees that, as in Section \ref{sec:higher}, the resulting model is equivalent to the pure CDT model (with staircase boundary conditions) up to a global rescaling of the coordinates, $\Lambda\!\mapsto\! \frac{\Lambda}{\Theta}$ and $L\!\mapsto\! \Theta L$ and hence belongs to the same universality class as the pure CDT model.

The above examples demonstrate that two-dimensional CDT define an entire universality class, implying that it is independent of the subtleties of the regularization and is a genuine continuum theory of Lorentzian quantum gravity, without evidence of a discretization of space-time in a physical sense.

\subsection{Causal dynamical triangulations as heaps of dimers and Dyck paths}\label{sec:dimer}

In this section we want to discuss bijections between CDTs, heaps of dimers and Dyck paths which were first established in \cite{DiFrancesco:1999em} in a slightly modified form. Such connections can be useful to reduce the two-dimensional counting problem to a one-dimensional combinatorial problem.

\begin{figure}[t]
\begin{center}
\includegraphics[width=5in]{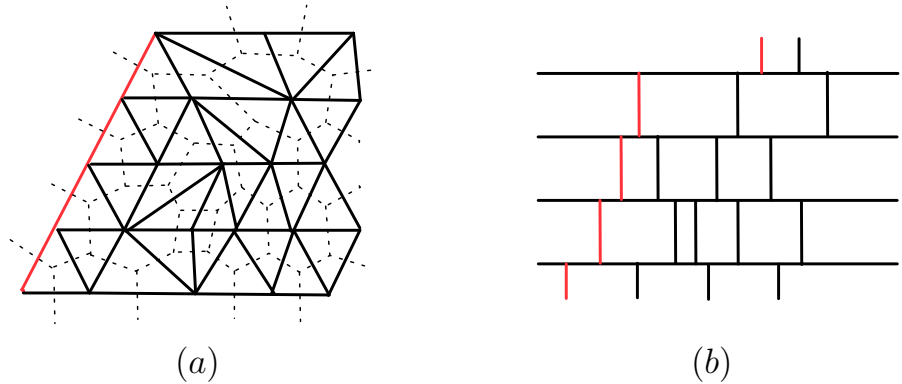}
\caption{(a) A triangulation of topology $[0,1]\!\times\![0,1]$ with all left-most triangles pointing upwards and no restrictions on the right-most triangles. Triangulations with these boundary conditions are in one-to-one correspondence with triangulations of topology $[0,1]\!\times\!S^1$ with a marking on the initial spatial boundary. (b) The corresponding dual triangulation.}\label{fig:dualleft}
\end{center}
\end{figure}

In the ``pure'' two-dimensional CDT model, as discussed in Section \ref{sec:CDT}, we considered space-time triangulations to have the topology $[0,1]\!\times\!S^1$ of a cylinder. Further, for combinatorial reasons, we introduced a marking on the initial spatial boundary. One can follow the right-most time-like edge connected to the marking to trace a unique path through the entire triangulation which can be described by a sequence of up-pointing triangles. Cutting open the geometry along this path one can establish a one-to-one correspondence between cylindrical triangulations with a marking on the initial spatial boundary and triangulations of topology $[0,1]\!\times\![0,1]$ whose left-most triangles are all pointing upwards, but without restrictions on the right-most triangles (cf. Appendix \ref{App:boundary}). A triangulation with such boundary conditions and the corresponding dual triangulation are illustrated in Figure \ref{fig:dualleft}.

\begin{figure}
\begin{center}
\includegraphics[width=6in]{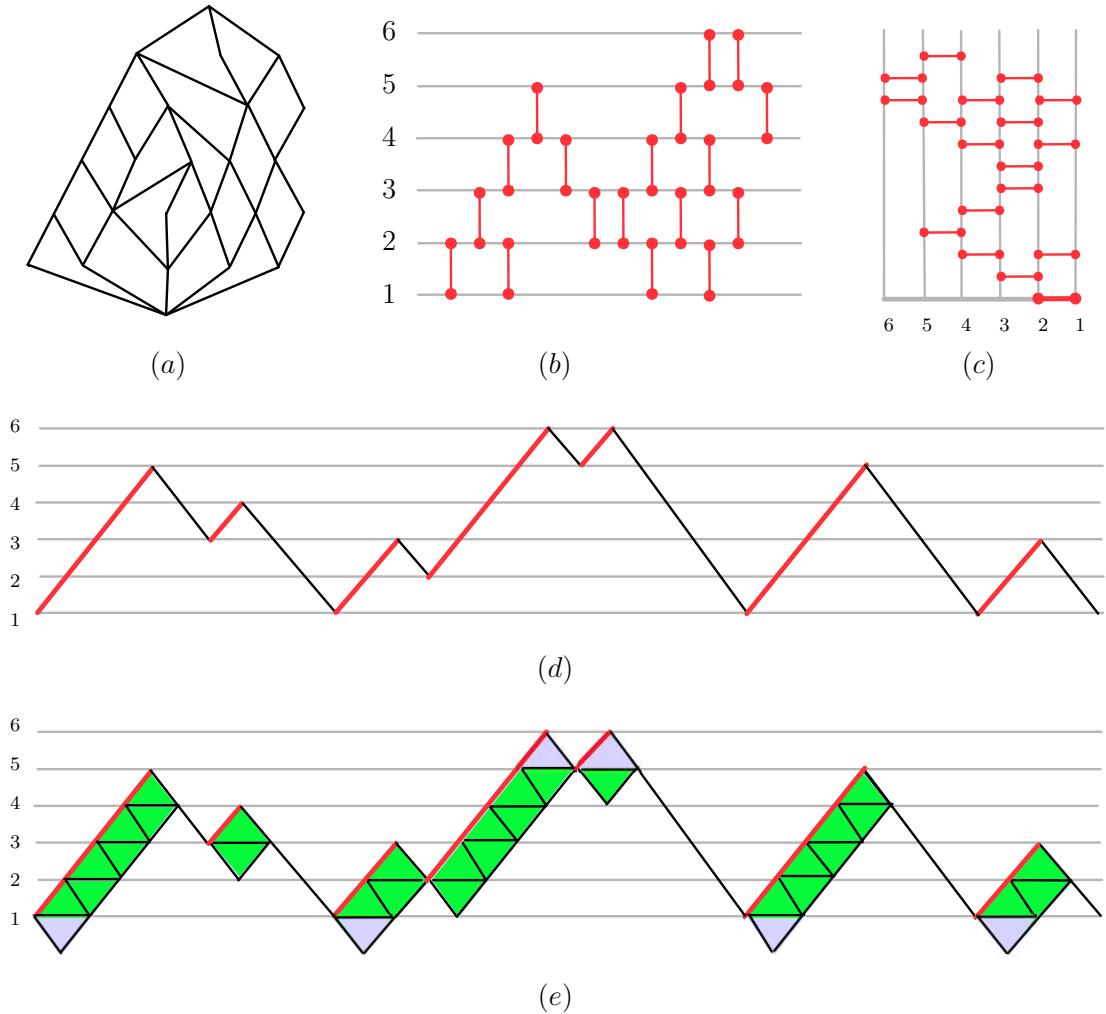}
\caption{(a) The tiling of diamonds corresponding to the triangulation shown in Figure \ref{fig:dualleft} (a) is obtained by extending the triangulation to a new vertex below and above and removing all space-like links. (b) Starting from the dual triangulation (Figure \ref{fig:dualleft} (b)) one obtains a half-pyramid of dimers by removing all space-like edges and pushing all remaining time-like edges (dimers) to the left until they touch each other. (c) Illustration of a half-pyramid of dimers with the base line on the bottom. One sees that only the right-most dimer is minimal. (d) Dyck path corresponding to the half-pyramid of dimers shown in (b). There is an one-to-one correspondence, where each up-wards step in the Dyck path corresponds to dimer in (b) with the respective ordering. (e) To obtain the original triangulation from the Dyck path one places a diamond (pair of triangles) under each upwards step of the Dyck path. Then, gluing together the triangles horizontally and removing the upper and lower row of triangles, one recovers the original triangulation.}\label{fig:tilingheap}
\end{center}
\end{figure}

Starting from a dual triangulation with this type of boundary conditions (Figure \ref{fig:dualleft} (b)) we can remove the space-like links and view the remaining time-like links as \textit{dimers}. Putting a base line to the left and pushing all dimers to the left until they touch the base line or another dimer we get a so-called \textit{heap of dimers}, as illustrated in Figure \ref{fig:tilingheap} (b). The dual of this heap of dimers is a tiling of diamonds (Figure \ref{fig:tilingheap} (a)), which one can obtain by extending the original triangulation to a vertex below and above and then removing all space-like links. The combinatorial structure of heaps of dimers was first introduced in \cite{heaps}. Usually heaps of dimers are illustrated with the base line on the bottom as shown in Figure \ref{fig:tilingheap} (c). If the orthogonal projection of the dimers onto the base line is connected the heap is \textit{connected}. Further, we call dimers touching the base line \textit{minimal} and heaps with just one minimal dimer are called \textit{pyramids}. If, moreover, this one minimal dimer lies in the right-most column, the heap is a \textit{half-pyramid}. It is not difficult to see that heaps of dimers corresponding to CDTs of all types of boundary conditions are connected. Moreover, for the type of boundary conditions considered above the corresponding heap of dimers is a half-pyramid, as shown in Figure \ref{fig:tilingheap} (c). This leads to a one-to-one correspondence between CDTs of topology $[0,1]\!\times\!S^1$ and half-pyramids of dimers. One can therefore enumerate CDTs and obtain the discrete finite-time propagator $G_\lambda(x,y;t)$ using the generating functions for connected heaps of dimers \cite{heaps2,james}. Hence, the solution of the discrete problem in two-dimensional CDT reduces to a one-dimensional counting problem.

There exists another interesting bijection between half-pyramids of dimers and Dyck paths. Consider a regular square lattice in the positive quadrant in the plane. A Dyck path is a one-dimensional random walk which starts at zero and finishes at zero without leaving the positive quadrant and only consisting of north-east steps $\nearrow$ and south-east steps $\searrow$ (Figure \ref{fig:tilingheap} (d)). 
The number of Dyck paths of length $2n$ is given by the n${}^{th}$ Catalan number $C_n$. 

The correspondence between half-pyramids of dimers and Dyck paths is illustrated in Figure \ref{fig:tilingheap} (d). Each dimer of the half-pyramid corresponds uniquely to an upwards step in the corresponding level of the Dyck path with the same relative ordering as in the half-pyramid.
This gives a one-to-one correspondence between half-pyramids of dimers and Dyck paths.

To obtain the original triangulation from the corresponding Dyck path one places a diamond (pair of triangles) under each upwards step in the Dyck path. Gluing together these triangles horizontally and removing the upper and lower row of triangles, one recovers the original triangulation.

In summary, we have shown the following bijections
\begin{equation}
\text{CDT}\leftrightarrow\text{half-pyramid of dimers}\leftrightarrow\text{Dyck path},
\end{equation}
which relate two-dimensional CDT to other well-known one-dimensional combinatorial structures.

\section{Conclusion}

In this review we have described the motivations and concepts behind the recent approach of Causal Dynamical Triangulation. In the case of two dimensions the model is analytically solvable as we demonstrated by explicit calculations. After the solution in the discrete setting was obtained, we showed the existence of a continuum limit, leading to a well-defined continuum theory of two-dimensional Lorentzian quantum gravity. We calculated several continuum quantities such as the effective quantum Hamiltonian, its spectrum and eigenfunctions, the finite time partition function and the loop-loop correlator. Using these results we obtained physical observables such as the expectation value of the length, higher moments and the effective Hausdorff dimension. In terms of these observables the resulting theory described a fluctuating two-dimensional quantum ``universe''. Further we have seen in this context that two-dimensional quantum gravity with Euclidean and Lorentzian signature are distinct theories.

In the second part of the review we discussed the possibility of including a sum over topologies in the path integral. The presence of causality constraints imposed on the
path-integral histories enabled us to derive a new class of continuum theories by taking an
unambiguously defined double-scaling limit of the bare cosmological and Newton's constant. These causality constraints were crucial to resolve the problem of super-exponential growth in the number of configurations which usually appears in the context of two-dimensional Euclidean quantum gravity with topology changes.  The resulting model of two-dimensional Lorentzian quantum gravity with topology changes has, besides the well-known geometrical observables, new ``topological'' observables, such as the finite space-time density of wormholes. Further, we observed that the presence of wormholes in our model leads to a decrease of the ``effective" cosmological constant, reminiscent of the suppression mechanism considered by Coleman and others in the context of a Euclidean path integral formulation of four-dimensional quantum gravity in the continuum.

In the last part we discussed aspects of universality of the two-dimensional model. We saw that including a higher curvature term in the action does not affect the continuum theory. Further, this higher curvature term could be interpreted as be equivalent to including squares in the space-time triangulation. This provided evidence that the same continuum theory can be obtained unambiguously without dependence on the details of the discrete building blocks.
Thus two-dimensional Lorentzian quantum gravity defined through CDT is a well-defined nonperturbative continuum theory.
 Further, we established certain one-to-one correspondences between CDTs, half-pyramids of dimers and Dyck paths. This provided a relation between the counting problem appearing in the discrete setting of CDT and certain well-known one-dimensional combinatorial structures.

In conclusion, we have seen that Causal Dynamical Triangulations is a promising nonperturbative approach to quantum gravity. It might be the right candidate for the challenging quest of quantizing gravity.

\section*{Acknowledgements}

The author thanks R. Loll for introducing him into the fascinating subject of Causal Dynamical Triangulations. Further, he would like to thank J. Ambj\o rn, J. Jers\'ak and W. Westra for enjoyable discussions. Support of the Dr.-Carl-Duisberg-Foundation
and a Theoretical Physics Utrecht Scholarship is gratefully acknowledged.

\appendix

\setcounter{equation}{0}
\renewcommand{\theequation}{A.\arabic{equation}}
\section{Lorentzian angles and simplicial building blocks}\label{App:Lorentzian}

In this appendix, a brief summary of results on Lorentzian angles is presented, where we follow the treatment and conventions of \cite{Sorkin:1975ah}.

\begin{figure}[t]
\begin{center}
\includegraphics[width=4.5in]{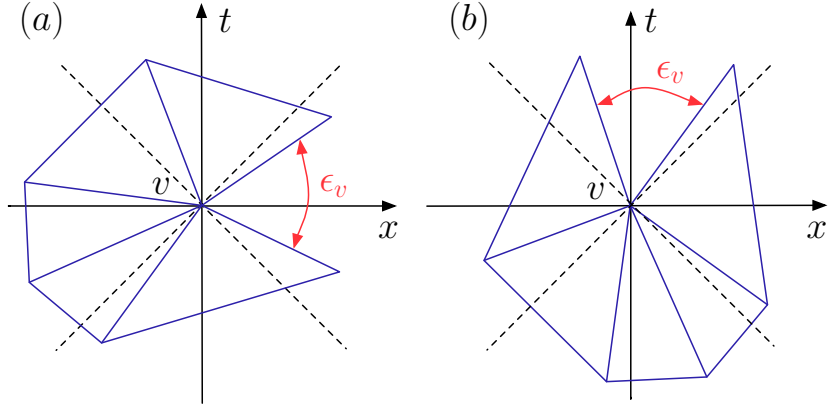}
\caption{Illustration of a space-like (a) and a time-like (b) Lorentzian deficit angle $\epsilon_v$ at a vertex $v$.}
\label{fig:appdeficit}
\end{center}
\end{figure}

Since in CDT one considers simplicial manifolds consisting of Minkowskian triangles,
Lorentzian angles or ``boosts'' naturally appear in the Regge action as rotations around vertices. Recall from Section \ref{sec:geometry} that the definition of the Gaussian curvature at a vertex $v$ is given by
\eqref{eq:simpl:gauss},
\begin{equation}\label{eq:applor:K}
K_v=\frac{\epsilon_v}{V_v},
\end{equation}
where $\epsilon_v=2\pi-\sum_{i\supset v}\theta_i$ is the deficit angle at a vertex $v$ and $V_v$ is the dual volume of the vertex $v$. Recall that the space-like deficit angle $\epsilon_v$ can be positive or negative as illustrated in Figure \ref{fig:curvature}. Furthermore, if the deficit angle is time-like, as shown in Figure \ref{fig:appdeficit}, it will be complex. The time-like deficit angles are still additive, but contribute to the curvature \eqref{eq:applor:K} with the opposite sign. Hence, both space-like defect and time-like excess increase the curvature, whereas space-like excess and time-like defect decrease it.
 
The complex nature of the time-like deficit angles can be seen explicitly by noting that the angles $\theta_i$ between two edges $\vec{a}_i$ and $\vec{b}_i$ (as vectors in Minkowski space) are calculated using
\begin{equation}
\label{eq:applor:theta}
\cos \theta_i =\frac{\scprod{\vec{a}_i}{\vec{b}_i}}{\scprod{\vec{a}_i}{\vec{a}_i}^\frac{1}{2}\scprod{\vec{b}_i}{\vec{b}_i}^\frac{1}{2}},\quad \sin \theta_i  
=\frac{\sqrt{\scprod{\vec{a}_i}{\vec{a}_i}\scprod{\vec{b}_i}{\vec{b}_i}-\scprod{\vec{a}_i}{\vec{b}_i}^2}}{\scprod{\vec{a}_i}{\vec{a}_i}^\frac{1}{2}\scprod{\vec{b}_i}{\vec{b}_i}^\frac{1}{2}},
\end{equation}
where $\scprod{\cdot}{\cdot}$ denotes the flat Minkowskian scalar product and by definition, the square roots of negative arguments are positive imaginary.

Having given a concrete meaning to  Lorentzian angles, we can now use   
\eqref{eq:applor:theta} to calculate the volume of Minkowskian triangles which we will then use to explicitly compute the volume terms of the Regge action. 

The triangulations we are considering consist of Minkowskian triangles with one space-like edge of length squared $l_s^2=a^2$ and two time-like edges of length squared $l_t^2=-\alpha a^2$ with $\alpha>0$. The general argument $\alpha>0$ is used to give a mathematically precise prescription of the Wick rotation, but it can be set to $\alpha=1$ after the Wick rotation has been performed. With the use of \eqref{eq:applor:theta} we can calculate the volume of such a Minkowskian triangle, yielding
\begin{equation}
\Vol(\mathrm{triangle})=\frac{a^2}{4}\sqrt{4\alpha+1}.
\end{equation}
Now one can define the Wick rotation $\mathcal{W}$ as the analytic continuation of $\alpha\mapsto-\alpha$ through the lower-half plane. One then sees that for $\alpha>\frac{1}{2}$ under this prescription $\mbox{$i\,\Vol(\mathrm{triangle})\mapsto -\Vol(\mathrm{triangle})$}$ (up to a $\mathcal{O}(1)$ constant which can be absorbed in the corresponding coupling constant in the action). This ensures that
\begin{equation}
\mathcal{W}:\quad  e^{i\,S_{\mathrm{Regge}}(T^{lor})}\mapsto  e^{-\,S_{\mathrm{Regge}}(T^{eu})}, \quad \alpha>\frac{1}{2}.
\end{equation}
In the following we set $\alpha=1$ again.
Generalizations of this treatment to dimension $d=3,4$ can be found in \cite{Ambjorn:2001cv}.

\setcounter{equation}{0}
\renewcommand{\theequation}{B.\arabic{equation}}
\section[Different boundary conditions]{Different boundary conditions and generating functions}\label{App:boundary}

In this section we want to investigate different spatial boundary conditions of the simplicial manifolds considered in the construction of CDT. As we have seen in Section \ref{sec:transfer}, the key to the solution of the model is knowing the eigenvalues of the transfer matrix,
\begin{equation}
\label{eq:appbound:onestepcomb}
G_\lambda(l_{in},l_{out};t=1)=e^{-\lambda\,a^2\,(l_{in}+l_{out})}\sum_{\substack{T\in \mathcal{W}(\Tri): \\ l_{in}\rightarrow l_{out}}} 1,
\end{equation} 
where all triangulations $T$ are single strips of height $\Delta t=1$. Considering a certain boundary condition on the simplicial manifold then reduces, in terms of the transfer matrix, to a simple combinatorial counting problem of evaluating the number of all possible configurations of a single strip with the same boundary conditions. The results for strips of different boundary conditions can be summarized in the following expression \cite{kappel}:
\begin{equation}
G_{\lambda,k}(l_{in},l_{out};t=1)=\frac{l_{in}^{k_{in}}\,
 l_{out}^{k_{out}}}{(l_{in}+l_{out})^{k_r}(l_{in}+l_{out}-1)^{k_{rr}}}\binom{l_{in}+l_{out}}{l_{in}} 
\,\,e^{-\lambda\,a^2\,(l_{in}+l_{out})},
\end{equation}
where $k$ denotes the multi index $k=(k_{in},k_{out},k_r,k_{rr})\in \{0,1\}^4$. Let us now discuss which cases of boundary condition we have considered:
\begin{enumerate}
\item Topology $[0,1]\!\times\! S^1$ (as already considered in Section \ref{sec:transfer}), corresponding to $k_r=1$ and $k_{rr}=0$. Furthermore, one can mark (not mark) the initial loop by setting $k_{in}=1$ ($k_{in}=0$) and respectively for $k_{out}$ for the final loop.
\item Topology $[0,1]\!\times\![0,1]$ without restrictions on the orientation of the left- and rightmost triangle, corresponding to $k_i=0$, for all $i\in\{in,out,r,rr\}$.
\item Topology $[0,1]\!\times\![0,1]$ without restriction of the triangle orientation on one side and with one up pointing triangle on the other side, corresponding to $k_{out},k_{rr}=0$ and $k_{in},k_{r}=1$.
\item Topology $[0,1]\!\times\![0,1]$ without restriction of the triangle orientation on one side and with one down pointing triangle on the other side, corresponding to $k_{in},k_{rr}=0$ and $k_{out},k_{r}=1$.  
\item Topology $[0,1]\!\times\![0,1]$ with one up pointing triangle on the one side and one down pointing triangle on the other side (``staircase''), corresponding to $k_i=1$, for all $i\in\{in,out,r,rr\}$.
\end{enumerate}

Another useful way of characterizing cases (1)-(5) is by their generating functions. Hereby one can take advantage of graphical methods, where one assigns a factor of $gx$ to every up pointing triangle ``$\bigtriangleup$'' and a factor of $gy$ to every down pointing triangle ``$\bigtriangledown$''. Considering the same cases as above:

\begin{enumerate}

  \item The generating function of the propagator for triangulations of topology $[0,1]\!\times\! S^1$ with one mark on the initial loop  we have already obtained in Section \ref{sec:transfer}, namely
  \begin{equation}
G^{\mathrm{circular}}_{\mathrm{in}}(x,y;g;1)=\frac{g^2xy}{(1-gx)(1-gx-gy)},
\end{equation}
and analogously for one mark on the final loop by interchanging $x$ and $y$.
The corresponding propagator with both loops marked then reads
  \begin{equation}
G^{\mathrm{circular}}_{\mathrm{in,out}}(x,y;g;1)=\frac{g^2xy}{(1-gx-gy)^2}.
\end{equation}
  
  \item The generating function of the propagator for triangulations of topology $[0,1]\times [0,1]$ without restrictions on either side is in most of the literature used in its ``degenerate'' form (where initial and final boundary of length zero are allowed),
\begin{equation}
G^{\mathrm{open}}(x,y;g;1)=1+\bigtriangleup+\bigtriangledown+...=\sum_{l=0}^\infty \left(\bigtriangleup +\bigtriangledown\right)^l=\frac{1}{1-gx-gy}.
\end{equation}

\item The case of open boundary on one side and one up pointing triangle on the other side can then easily be generated by
\begin{equation}
G^{\mathrm{open}}_{\bigtriangleup}(x,y;g;1)=\bigtriangleup\cdot G^{\mathrm{open}}(x,y;g;1)=\frac{gx}{1-gx-gy},
\end{equation}
which includes the zero loop in the final spatial boundary.

\item The case of open boundary on one side and one down pointing triangle on the other side is respectively generated by
\begin{equation}
G^{\mathrm{open}}_{\bigtriangledown}(x,y;g;1)=\bigtriangledown\cdot G^{\mathrm{open}}(x,y;g;1)=\frac{gy}{1-gx-gy},
\end{equation}
which includes the zero loop in the initial spatial boundary.

\item The generating function of the propagator for triangulations of topology $[0,1]\times [0,1]$ with one up pointing triangle on one side and one up pointing triangle on the other side is given by
\begin{equation}
\label{eq:appbound:staircase}
G^{\mathrm{open}}_{\mathrm{staircase}}(x,y;g;1)=\bigtriangleup\cdot G^{\mathrm{open}}(x,y;g;1)\cdot\bigtriangledown=\frac{g^2xy}{1-gx-gy},
\end{equation}
which has obviously no degeneracies.
\end{enumerate}

Triangulations with staircase boundaries can be related to triangulations of topology $[0,1]\!\times\! S^1$ with both spatial boundaries marked \cite{DiFrancesco:1999em}. 

\begin{figure}[t]
\begin{center}
\includegraphics[width=6in]{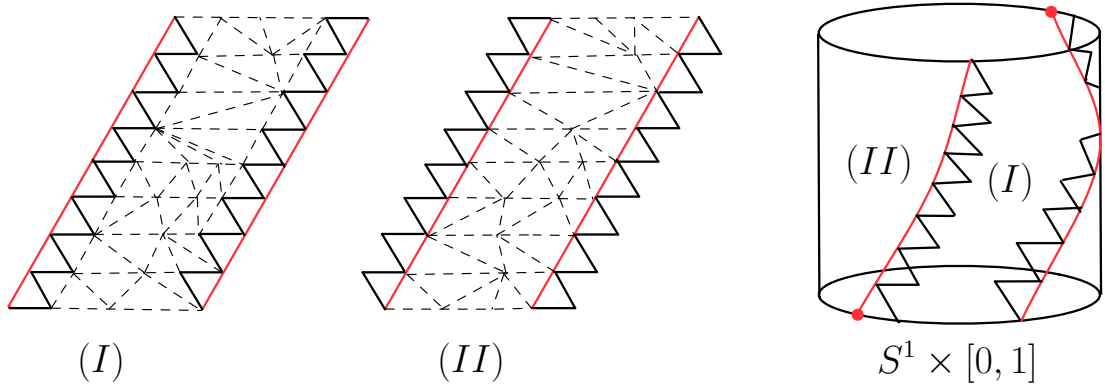}
\caption{Construction of a triangulation of topology $[0,1]\!\times\! S^1$ with both spatial boundaries marked by a composition of two triangulations with staircase boundary (I) and (II).}
\label{fig:apptwoseemed}
\end{center}
\end{figure}

Therefore consider two triangulations with staircase boundary, one which has the sequence of up pointing triangle on the left (I) and one which has the sequence of down pointing triangles on the left (II) (Figure \ref{fig:apptwoseemed}). The triangulation of topology $[0,1]\!\times\! S^1$ with both spatial boundaries marked can then be obtained by gluing triangulations (I) and (II) to a cylinder, where one superimposes the respective left and right most triangles. Both of the sequences of up and down pointing triangles uniquely define both of the markings on the initial and final spatial boundary (Figure \ref{fig:apptwoseemed}). In terms of generating functions of the one-step propagator, the correspondence can easily be seen by noting that the triangulations (I) and (II) both have the same one-step propagator \eqref{eq:appbound:staircase}. The one-step propagator of the composition reads
\begin{eqnarray}
G^{\mathrm{circular}}_{\mathrm{in,out}}(x,y;g;1) & = & \frac{1}{g^2 x y} \left( G^{\mathrm{open}}_{\mathrm{staircase}}(x,y;g;1) \right)^2 \nonumber\\
 & = & \frac{g^2xy}{(1-gx-gy)^2}, \label{eq:appbound:twoseamed}
\end{eqnarray}
where the division by $g^2xy$ is due to the overlap of the left and right most triangles of (I) and (II). The composition \eqref{eq:appbound:twoseamed} is sometimes denoted as two-seamed correlator. A generalization of this treatment, where one considers more than two strips (of either type (I) or (II)) in the construction of a cylindrical triangulation can be found in \cite{DiFrancesco:1999em}.

\setcounter{equation}{0}
\renewcommand{\theequation}{C.\arabic{equation}}
\section{An alternative derivation of the Hamiltonian}\label{App:hamilton}

In this appendix an alternative derivation of the Hamiltonian is presented, where in contrast to Section \ref{sec:cont}, the continuum limit is obtained in length space instead of the Laplace transformed ``momentum space''.

Starting point of the derivation is the one-step propagator for general boundary conditions (in the Euclidean sector), as presented in Appendix \ref{App:boundary},
\begin{equation}
\label{eq:appdiff:prop}
G_{\lambda,k}(l_{1},l_{2};t=1)=\bra{l_2}\Trans \ket{l_1}=\frac{l_{1}^{k_{in}}\,
 l_{2}^{k_{out}}}{(l_{1}+l_{2})^{k_r}(l_{1}+l_{2}-1)^{k_{rr}}}\binom{l_{1}+l_{2}}{l_{1}} 
\,\,e^{-\lambda\,a^2\,(l_{1}+l_{2})}.
\end{equation}
To obtain the continuum limit we use the following canonical scaling relations
\begin{equation}
\label{eq:appdiff:scaling1}
L_1=l_1\cdot a,\quad L_2=l_2\cdot a,\quad T=t\cdot a,
\end{equation}
where the physical correlation lengths $L_1$, $L_2$ and $T$ stay fixed under the simultaneous limit $a\rightarrow 0$ and $l_1,l_2,t\rightarrow\infty$. Further, one has to introduce an additive renormalization for the bare cosmological constant $\lambda$ (cf. Section \ref{sec:cont})
\begin{equation}
\label{eq:appdiff:scaling2}
\lambda=\frac{\log 2}{a^2}+\Lambda,		
\end{equation} 
where $\Lambda$ denotes the renormalized cosmological constant. The continuum propagator can then be obtained by
\begin{equation}
\label{eq:appdiff:scaling3}
G_{\Lambda,k}(L_{1},L_{2};T)\xrightarrow[]{a\rightarrow 0} \,a^\eta\, G_{\lambda=\frac{\log 2}{a^2}+\Lambda,k}\left(l_{1}=\frac{L_1}{a},l_{2}=\frac{L_2}{a};t=\frac{T}{a}\right),
\end{equation}
where $a^\eta$ is a wave function renormalization. To extract the Hamiltonian we consider the infinitesimal time evolution of the wave function,
\begin{eqnarray}
\label{eq:appdiff:timeevl}
\psi(L_2,T+a)&=&\int_0^\infty d\mu(L_1)\,G_{\Lambda,k}(L_{1},L_{2};T=a) \,\psi(L_1,T)\nonumber\\
&=&\int_0^\infty d\mu(L_1)\,\bra{L_2}(1-a\,\hat{H}+\mathcal{O}(a^2))\ket{L_1}\, \psi(L_1,T),
\end{eqnarray}
with the measure $d\mu(L_1)\equ L_1^\mu dL_1$. For the further calculation it will be useful to introduce a new length variable $D\!:=\! L_2\mi L_1$. In terms of this variable we can rewrite \eqref{eq:appdiff:timeevl} as
\begin{equation}
\label{eq:appdiff:timeev}
\psi(L_2,T+a)=\int_{-\infty}^{L_2} dD (L_2-D)^\mu\, G_{\Lambda,k}(L_{2}-D,L_{2};T=a) \,\psi(L_2-D,T).
\end{equation}
In the following we want to determine $G_{\Lambda,k}(L_{2}\mi D,L_{2};T\equ a)$ up to first order in $a$. Therefore we start expanding \eqref{eq:appdiff:prop} for large $l_i$ and introduce new variables $l\! :=\! l_1\pl l_1$ and $d\! :=\! l_2\mi l_1$. We first consider the binomial factor in \eqref{eq:appdiff:prop} making use of Stirling's approximation $n! \equ \sqrt{2\pi n}\left(\frac{n}{e}\right)^n\left(1+\frac{1}{12n}+\mathcal{O}(\frac{1}{n^2})\right)$ \cite{abramowitz}, hence
\begin{eqnarray}
A\!\!\!&:=&\!\!\!\!\binom{l_{1}+l_{2}}{l_{1}} 
\,\,e^{-\lambda\,a^2\,(l_{1}+l_{2})}  =  \frac{l!}{\left(\frac{l}{2}\left(1-\frac{d}{l}\right)\right)!\left(\frac{l}{2}\left(1+\frac{d}{l}\right)\right)!}\,\, e^{-a^2\lambda\, l} \nonumber\\
  &=& \!\!\!\! e^{-a^2\lambda\, l}\sqrt{\frac{2}{\pi l}}\,2^l(1-\frac{d^2}{l^2})^{-\frac{1}{2}}\!\left(1+\frac{1}{12l}-\frac{1}{6l(1-\frac{d}{l})}-\frac{1}{6l(1+\frac{d}{l})}+\mathcal{O}(\frac{1}{l^2})\right)\!\! F(l,d),\quad\quad \label{eq:appdiff:binom}    
\end{eqnarray}
with
\begin{equation}
\label{eq:appdiff:F}
F(l,d)=\left(1-\frac{d}{l}\right)^{-\frac{l}{2}\left(1-\frac{d}{l}\right)}\left(1+\frac{d}{l}\right)^{-\frac{l}{2}\left(1+\frac{d}{l}\right)}=e^{-\frac{l}{2}\left(1-\frac{d}{l}\right)\log\left(1-\frac{d}{l}\right)
-\frac{l}{2}\left(1+\frac{d}{l}\right)\log\left(1+\frac{d}{l}\right)}.
\end{equation}
Introducing the scaling relations \eqref{eq:appdiff:scaling1} into \eqref{eq:appdiff:F}, one can see that to lowest order it behaves like $F(L_2,D)\!\sim\! \exp( -\frac{D^2}{4aL_2})$. Since we want to obtain a delta function to lowest order, i.e. $F(L_2,D)\!\sim\!\delta(D)$, we can conclude that $D$ must scale as $D\!\sim\!\sqrt{a}$ as $a\rightarrow 0$. Keeping this in mind, we can introduce the scaling relations \eqref{eq:appdiff:scaling1}-\eqref{eq:appdiff:scaling2} into \eqref{eq:appdiff:binom} and collect all terms up to first order in $a$, yielding
\begin{eqnarray}
A=\sqrt{\frac{a}{\pi L_2}}e^{-\frac{D^2}{4aL_2}}\!\!\!\! &\!&\! \left(
1+\frac{D}{4L_2}-\frac{D^3}{8aL_2}
-2a\Lambda L_2\right.-
\nonumber\\
&-&\left. \frac{a}{8L_2}+\frac{7}{32}\frac{D^2}{L_2^2}-\frac{10}{96}\frac{D^4}{aL_2^3}+\frac{D^6}{128a^2L_2^4}+\mathcal{O}(a^\frac{3}{2})\right)
\end{eqnarray}
In an analogous way one can also expand the remaining part of the propagator including the measure factor and the wave function renormalization factor,
\begin{eqnarray}
B & := & a^\eta\,l_1^{k_{in}+\mu}l_2^{k_{out}}(l_1+l_2)^{-k_r}(l_1+l_2-1)^{-k_{rr}} \nonumber\\
 & = & 2^{-k_r-k_{rr}}a^{\eta}\left(\frac{L_2}{a}\right)^{\mu+k_{in}+k_{out}-k_r-k_{rr}}
\left\{ 1+ \frac{D}{2L_2}(k_r+k_{rr}-2k_{in}-2\mu)+\right.
\nonumber\\
&+& k_{rr}\frac{a}{2L_2}+\frac{D^2}{8L_2^2}[ 2(k_r+k_{rr}+k_r k_{rr})+4(k_{in}+\mu)(k_{in}+\mu-1)-\nonumber\\
&-& \left.4(k_r+k_{rr})(k_{in}+\mu) ]+\mathcal{O}(a^\frac{3}{2})\right\}, 
\end{eqnarray}
where we used that $k_i\equ k_i^2$. Having obtained the continuum propagator up to first order in $a$, we can now write \eqref{eq:appdiff:timeev} in the limit $a\rightarrow 0$ as
\begin{equation}\label{eq:appdiff:timefinal}
\!\!\psi(L_2,T+a)=\int_{-\infty}^{\infty}\!\!\!\! dD A(L_2,D)\,B(L_2,D) \!\!\left(1\mi D\deriv{}{L_2}\pl\frac{D^2}{2}\deriv{{}^2}{{L_2}^2}\pl \mathcal{O}(a^\frac{3}{2}) \right)\psi(L_2,T).
\end{equation}
Upon introducing $A(L_2,D)$ and $B(L_2,D)$ as obtained above, one can then simply perform the gaussian integrals in \eqref{eq:appdiff:timefinal} by doing the variable substitution $D\mapsto\tilde{D}\equ D/\sqrt{2 a L_2}$. The final result reads
\begin{eqnarray}
\label{eq:appdiff:timeH}
\psi(L_2,T+a)=2^{1-k_r-k_{rr}}a^{1+\eta}\left(\frac{L_2}{a}\right)^{\mu+k_{in}+k_{out}-k_r-k_{rr}}
\!\!\!\left(1-a\hat{H}+\mathcal{O}(a^\frac{3}{2})\right)\psi(L_2,T),
\end{eqnarray}
where the Hamiltonian is given by
\begin{equation}
\label{eq:appdiff:H}
\hat{H}(L_2,\deriv{}{L_2})=-L_2\deriv{{}^2}{L_2^2}-c_1 \deriv{}{L_2} -c_2\frac{1}{L_2} +2\,\Lambda\,L_2
\end{equation} 
with the coefficients
\begin{eqnarray}
c_1 & = & 1+2(k_{in}+\mu)-k_r-k_{rr} \\
c_2 & = & (k_{in}+\mu)^2-(k_r+k_{rr})(k_{in}+\mu)+\frac{1}{2}k_r k_{rr}+\frac{1}{2}k_{rr}. 
\end{eqnarray}
From \eqref{eq:appdiff:timeH} one can further see that to obtain the right behavior to lowest order we must have $\eta\equ -1$ and $\mu\equ k_r\pl k_{rr}\mi k_{in}\mi k_{out}$ (the remaining constant factor can be absorbed into the measure).
For the Hamiltonian \eqref{eq:appdiff:H} to be selfadjoint on the Hilbert space $\mathcal{H}\equ \mathcal{L}^2(\mathbb{R}_+,L_2^\mu dL_2)$ with $\mu\equ k_r\pl k_{rr}\mi k_{in}\mi k_{out}$, we must symmetrize the Hamiltonian by setting $k_{in}\equ k_{out}$.

Consider for example the case of topology $[0,1]\!\times\! S^1$, corresponding to $k_{rr}\equ 0$ and $k_r\equ 1$; formula \eqref{eq:appdiff:timeH} then reduces to the Hamiltonian for the case with one marking on the initial spatial boundary, i.e. $k_{in}\equ 1$,
\begin{equation}
\hat{H}(L,\deriv{}{L})=-L\ddL+2\,\Lambda\,L,\quad d\mu(L)=L^{-1}dL\quad\text{(top. } [0,1]\times S^1\text{, marked)}
\end{equation}
and the unmarked case, $k_{in}\equ0$,
\begin{equation}
\hat{H}(L,\deriv{}{L})=-L\ddL -2\dL +2\,\Lambda\,L,\quad d\mu(L)=LdL\quad\text{(top. } [0,1] \times S^1\text{, unmarked)},
\end{equation}
which coincides with the results obtained in Section \ref{sec:cont}. Another Hamiltonian one encounters frequently is the one belonging to topology $[0,1]\!\times\! [0,1]$ without restrictions on the time-like boundaries, i.e. $k_i\equ 0$,
\begin{equation}
\hat{H}(L,\deriv{}{L})=-L\ddL -\dL +2\,\Lambda\,L,\quad d\mu(L)=dL\quad\text{(top. } [0,1]\times [0,1]\text{)},
\end{equation}  
One sees that the Hamiltonian for open boundary conditions is very similar to the one for cylindrical boundary conditions. Further, the spectra are the same up to some difference in the zero point energy (cf. Appendix \ref{App:Calogero}). Therefore, one often uses the open boundary conditions instead of cylindrical boundary conditions for simplicity (as we partly do in Section \ref{sec:Topo}).

\setcounter{equation}{0}
\renewcommand{\theequation}{D.\arabic{equation}}
\section{Generalizations and the Calogero Hamiltonian}\label{App:Calogero}

In this appendix we want to analyze the properties and spectrum of Hamiltonians which one usually encounters when studying two-dimensional Lorentzian quantum gravity. As shown in \cite{DiFrancesco:2000nn}, these models can be related to the well-known Calogero model. 

In the previous appendix we have already seen that for different boundary conditions and different topology of the spatial slices the Hamiltonians differ slightly, but are generally of the form
\begin{equation}\label{calogero1}
\hat{H}(L,\dL)=-c_1 L\ddL-c_2 \dL+2 \Lambda L
\end{equation}
where $c_1$ and $c_2$ are real constants. In the following we therefore want to investigate this generalized class of Hamiltonians. 

It is immediately checked that the Hamiltonian \eqref{calogero1} is selfadjoint on the Hilbert space $\mathcal{H}\equ \mathcal{L}^2(\mathbb{R}_+,d\mu(L))$, where the measure $d\mu(L)$ is given by\footnote{Note that in \cite{DiFrancesco:2000nn} the definition of $\mu$ differs by one to the one used here.} 
\begin{equation}
d\mu(L)=L^\mu dL,\quad\mu=\frac{c_2}{c_1}-1.
\end{equation} 
Further, for the boundary components of the partial integration in $\braket{\hat{H}\psi}{\phi}\equ\braket{\psi}{\hat{H}\phi}$ to vanish, the wave functions on the Hilbert space must satisfy
\begin{equation}
\label{calogerowave}
\left[ L^{\mu+1}\psi(L) \dL\psi(L) \right]_0^\infty=0
\end{equation}
 
To have a Hilbert space with flat measure, one can pull the measure into the wave function by substituting $\Psi(L)\equ L^{-\frac{\mu}{2}}\psi(L)$, where $\psi(L)$ is the wave function corresponding to \eqref{calogero1}. Commuting  $L^{-\frac{\mu}{2}}$ through the Hamiltonian gives
\begin{equation}\label{calogero3}
\hat{H}^{flat}(L,\dL)=-c_1 L\ddL-c_1\dL+2\Lambda L +\frac{(c_1-c_2)^2}{4c_1}\frac{1}{L},\quad d\mu(L)=dL
\end{equation}
This pulling in and out of the measure is similar to what one does when introducing markings on the boundary loops.

We now consider the eigenvalue problem of the Hamiltonian \eqref{calogero3},
 \begin{equation}\label{calogero4}
\left(-c_1 L\ddL-c_1\dL+2\Lambda L +\frac{(c_1-c_2)^2}{4c_1}\frac{1}{L}-E\right)\Psi(L)=0.
\end{equation}
Let us perform a change of variables and wave functions,
 \begin{equation}
L=\frac{c_1}{2}\vph^2, \quad \Phi(\vph)=\sqrt{\vph} \Psi\left(\frac{\vph^2}{2}\right),
\end{equation}
where the latter guarantees a flat measure $d\mu(\vph)\equ d\vph$ for $\Phi(\vph)$. The eigenvalue problem then reads
 \begin{equation}\label{calogero6}
\left(-\frac{1}{2}\ddphi+\frac{1}{2}\omega^2\vph^2-\frac{1}{8}\frac{A}{\vph^2}-E\right)\Phi(\vph)=0,
\end{equation}
where we set $\omega\equ \sqrt{2c_1\Lambda}$ and $A\equ 1\mi 4\mu^2\!\in\!(-\infty,1)$. This is nothing but the eigenvalue problem corresponding to the Hamiltonian of the one-dimensional Calogero model,
\begin{equation}
\hat{H}^{Calogero}(\vph,\dphi)=-\frac{1}{2}\ddphi+\frac{1}{2}\omega^2\vph^2-\frac{1}{8}\frac{A}{\vph^2},\quad d\mu(\vph)=d\vph.
\end{equation}
Note that the parameter range $A\equ 1\mi 4\mu^2\!\in\!(-\infty,1)$ is the maximum range for which the Calogero Hamiltonian is selfadjoint. Two-dimensional Lorentzian quantum gravity with open boundary conditions corresponds to the value $A=1$, whereas for circular boundary conditions it corresponds to $A\equ -3$. In Appendix \ref{app:scaling} an integrable model is introduced which continuously covers the parameter range $-3\!\leqslant\! A\!\leqslant\! 1$. The first connection between 2d CDT and the Calogero model was established in \cite{DiFrancesco:2000nn}, where a generalized CDT model was introduced which continuously covered the parameter range $0\!\leqslant\! A\!\leqslant\! 1$.

The spectrum of the Calogero Hamiltonian is well-known, further it can be related to the spectrum of the radial solution of the three-dimensional Schr\"odinger equation with potential $U(r)=D_1/r^2+D_2 r^2$ \cite{landau}. Explicitly, the solution of the eigenvalue problem \eqref{calogero6} can be obtained by doing a variable transformation $\tilde{L}\equ\omega\vph^2$, yielding
\begin{equation}\label{calogero8}
\left(\tilde{L}\frac{\partial^2}{\partial\tilde{L}^2}+\frac{1}{2}\frac{\partial}{\partial\tilde{L}}-\frac{1}{4}\tilde{L}+\frac{E}{2\omega}+\frac{A}{16\tilde{L}}\right)\Phi(\tilde{L})=0.
\end{equation}
Note that $\tilde{L}$ is just a scaled version of $L$, so we could have directly arrived here from \eqref{calogero4}, using the wave function transformation from $\Psi(L)$ to $\Phi(\tilde{L})$. From \eqref{calogero8} one can see that asymptotically for $\tilde{L}\rightarrow\infty$ the solution should be proportional to $\exp(-\tilde{L}/2)$ and for small $\tilde{L}\rightarrow 0$ it behaves like $\tilde{L}^{1/4+\mu/2}$. Note that these are the only asymptotics compatible with the requirement \eqref{calogerowave} on the wave functions.  Hence, we make the following $ansatz$ for the wave functions $\Phi(\tilde{L})\propto\exp(-\tilde{L}/2)\tilde{L}^{1/4+\mu/2}\eta(\tilde{L})$. Substituting this into \eqref{calogero8} yields
 \begin{equation}\label{calogero9}
\tilde{L}\eta''(\tilde{L})+(1+\mu-\tilde{L})\eta'(\tilde{L})+\left(\frac{E}{2\omega}-\frac{1}{2}(\mu+1)\right)\eta(\tilde{L})=0
\end{equation}
This equation is known as Kummer's equation \cite{abramowitz}, whose solutions are the confluent hypergeometric functions,
\begin{equation}\label{calogero10}
\eta(\tilde{L})= \, {}_1 F_1 (-n;1+\mu;\tilde{L}),
\end{equation}
where $n\equ\frac{E}{2\omega}-\frac{1}{2}(1+\mu)$ must be a nonnegative integer. In this case the power series of the confluent hypergeometric function is truncated to a polynomial of degree $n$, namely, the generalized Laguerre polynomials \cite{gradshteyn},  
\begin{equation}
{}_1F_{1}(-n;\mu+1;z)=\frac{\Gamma(n+1)\Gamma(\mu+1)}{\Gamma(n+\mu+1)}\,L_n^\mu(z),
\end{equation}
with
\begin{eqnarray}
L_n^\mu(z)& = & \frac{1}{n!}e^z z^{-\mu} \frac{d^n}{dz^n}\left( e^{-z}z^{n+\mu} \right) \\
 & = & \sum_{k=0}^n (-1)^k \frac{\Gamma(n+\mu+1)}{\Gamma(n-k+1)\Gamma(\mu+k+1)}\frac{z^k}{k!},
\end{eqnarray}
yielding the final expression,
\begin{equation}
{}_1F_{1}(-n;\mu+1;z) =\sum_{k=0}^n (-1)^k \binom{n}{k}\frac{1}{(\mu+1)(\mu+2)...(\mu+k)}\frac{z^k}{k!}.
\end{equation}
Since $n\equ\frac{E}{2\omega}\mi\frac{1}{2}(1\pl\mu)$ must be a nonnegative integer, the spectrum of the Calogero Hamiltonian $\hat{H}^{Calogero}$ reads
\begin{equation}\label{calogero11}
E_n=\omega(2n+\mu+1),\quad n=0,1,2,...
\end{equation}
The corresponding eigenfunctions are given by
\begin{equation}\label{calogero:wavecalo}
	\Phi_n(\vph)=\mathcal{C}_n e^{-\frac{\omega}{2}\vph^2}\vph^{\frac{1}{2}+\mu}
	\,{}_1F_{1}(-n;1+\mu;\omega\vph^2),\quad d\mu(\vph)=d\vph.
\end{equation}
Further, since the Laguerre polynomials are orthogonal functions, the wave functions form an orthonormal basis of the Hilbert space, where the normalization constant is given by
\begin{equation}
\mathcal{C}_n=\sqrt{\frac{2\omega^{\mu+1}\Gamma(n+\mu+1)}{\Gamma(n+1)\Gamma(\mu+1)^2}},
\end{equation}
where we used the following orthogonal relation for the generalized Laguerre polynomials \cite{gradshteyn},
\begin{eqnarray}
\int_0^\infty d\tilde{L} e^{-\tilde{L}}\tilde{L}^\mu \, \mathrm{ L}_n^\mu(\tilde{L})  \mathrm{ L}_m^\mu(\tilde{L})=\begin{cases}
       &0 \quad \quad\quad\quad \text{for } n\neq m, \\
       &\frac{\Gamma(\mu+n+1)}{\Gamma(n+1)} 
     \quad\text{for } n=m,
\end{cases} 
\end{eqnarray}
which is valid for $\mu\!\geqslant\! 0$
.

Let us now come back to the original problem, analyzing the Hamiltonian
\begin{equation}
\hat{H}(L,\dL)=-c_1 L\ddL-c_2\dL+2 \Lambda L,\quad d\mu(L)=L^\mu dL,\quad\mu=\frac{c_2}{c_1}-1.
\end{equation}
The spectrum is obviously the same as for the Calogero Hamiltonian, hence from \eqref{calogero11} we get
\begin{equation}\label{calogeroenergy}
E_n=\sqrt{2c_1\Lambda}(2n+\mu+1),\quad n=0,1,2,...
\end{equation}
Further, the eigenfunctions of $\hat{H}(L,\partial_L)$ can be obtained from \eqref{calogero:wavecalo} by a simultaneous variable and wave function transformation, yielding
\begin{equation}\label{calogero15}
	\psi_n(L)= \mathcal{A}_n 
	e^{-\sqrt{\frac{2\Lambda}{c_1}}L}\,
	{}_1F_1(-n;1+\mu;2\sqrt{\frac{2\Lambda}{c_1}}L),\quad d\mu(L)=L^\mu dL
\end{equation}
where the normalization factor is given by
\begin{equation}
\mathcal{A}_n=2^\frac{\mu}{2} c_1^{-\frac{\mu+1}{2}} \mathcal{C}_n=
\left(\frac{8\Lambda}{c_1}\right)^{\frac{\mu+1}{4}}\sqrt{\frac{\Gamma(n+\mu+1)}{\Gamma(n+1)\Gamma(\mu+1)^2}}.
\end{equation} 

Having obtained the eigenfunctions for the class of Hamiltonians $\hat{H}(L,\partial_L)$, we are now able to give an explicit expression for the (Euclidean) finite time propagator or loop-loop correlator, as defined in Section \ref{sec:cont},
\begin{equation}
	G_\Lambda(L_1,L_2;T)=\bra{L_2}e^{-T\hat{H}}\ket{L_1}.
\end{equation}
Since the eigenfunctions form an orthonormal basis of the Hilbert space, we can insert the unit operator $I\equ \sum_{n \geqslant0}\ket{n}\bra{n}$, yielding
\begin{eqnarray}
	G_\Lambda(L_1,L_2;T) &=&\sum_{n=0}^\infty \braket{L_2}{n}
	e^{-TE_n}\braket{n}{L_1}\nonumber\\
	&=& \sum_{n=0}^\infty e^{-TE_n}\psi_n^*(L_2)\psi_n(L_1)\label{calogero:propdef}
\end{eqnarray}
Upon inserting the energy eigenvalues \eqref{calogeroenergy} and eigenfunctions \eqref{calogero15} into \eqref{calogero:propdef} one obtains
\begin{eqnarray}
	G_\Lambda(L_1,L_2;T)=\left(\frac{8\Lambda}{c_1}\right)^{\frac{\mu+1}{2}}
	e^{-\sqrt{\frac{2\Lambda}{c_1}}} \sum_{n=0}^\infty  z^{n+\frac{\mu+1}{2}}
	\frac{\Gamma(n+\mu+1)}{\Gamma(n+1)\Gamma(\mu+1)^2} \times\nonumber\\ 
	\times {}_1F_1(-n;1+\mu;2\sqrt{\frac{2\Lambda}{c_1}}L_1)\,{}_1F_1(-n;1+\mu;2\sqrt{\frac{2\Lambda}{c_1}}L_2),
\end{eqnarray}
where we used the notation $z\equ e^{-2\sqrt{2c_1\Lambda}T}$. To evaluate the above summation we make use of the following quadratic relation satisfied by the confluent hypergeometric function \cite{gradshteyn},
\begin{eqnarray}
	\sum_{n=0}^\infty  z^{n}
	\frac{\Gamma(n+\mu+1)}{\Gamma(n+1)\Gamma(\mu+1)^2}\, {}_1F_1(-n;1+\mu;x)\, {}_1F_1(-n;1+\mu;y)=\nonumber\\
	=\frac{1}{1-z}e^{-\frac{z(x+y)}{1-z}}(xyz)^{\frac{-\mu}{2}}I_{\mu}\left(2\frac{\sqrt{xyz}}{1-z}\right),
\end{eqnarray}
where $I_\mu(x)$ denotes the modified Bessel function of the $\mu$-th kind. Gathering all terms together leads to the final expression for the (Euclidean) finite time propagator
\begin{eqnarray}\label{calogero20}
	G_\Lambda(L_1,L_2,T)=\sqrt{\frac{\Lambda}{c_1}}(L_1L_2)^{-\frac{\mu}{2}}
	\frac{e^{-\sqrt{\frac{\Lambda}{c_1}}(L_1+L_2)\coth(\sqrt{c_1\Lambda}T)}}{\sinh(\sqrt{c_1\Lambda}T)}
	I_\mu\left(\frac{2}{\sqrt{c_1}}\frac{\sqrt{\Lambda L_1 L_2}}{\sinh(\sqrt{c_1\Lambda}T)}\right).
\end{eqnarray}
For $\mu\equ 1$, \eqref{calogero20} corresponds to the ``unmarked'' propagator for the 2d CDT model with circular boundary conditions, as obtained in \eqref{eq:cont:propagatorres}, where the identity was given by
\begin{equation}\label{eq:calogerocomplaw}
\braket{L_1}{L_2}=\frac{1}{L_1}\delta(L_1-L_2),\quad \int_0^\infty dL\ket{L} L \bra{L}=\hat{1}
\end{equation}
Note that by changing the measure one can always shift around factors of $L$'s from \eqref{calogero20} to \eqref{eq:calogerocomplaw}.
\\

In Section \ref{sec:obs} we were interested in calculating expectation values of the length and higher moments with respect to a certain eigenstate $\ket{n}$,
\begin{equation}
\label{calogeromoments}
\expec{L^m}_n\equiv\bra{n}L^m\ket{n}=\int_0^\infty d\mu(L) \psi^*_n(L)\,L^m\,\psi_n(L). 
\end{equation}
The result can also be obtained for the generalized class of Hamiltonians \eqref{calogero1}, as considered above. Inserting the eigenfunctions \eqref{calogero15} into \eqref{calogeromoments} yields
\begin{equation}
\label{calogeromoments2}
\expec{L^m}_n=\left(\frac{c_1}{4\Lambda}\right)^{\frac{m}{2}}\frac{\Gamma(n+1)}{\Gamma(n+\mu+1)} \int_0^\infty d\tilde{L}\, \tilde{L}^{\mu+m}\, e^{-\tilde{L}}  \, \left[\mathrm{ L}_n^\mu(\tilde{L})\right]^2.
\end{equation}
For the evaluation of the integral we use the following integral expression \cite{Prudnikov},
\begin{eqnarray}
\int_0^\infty d\tilde{L}\, \tilde{L}^{\mu+m}\, e^{-\tilde{L}}  \, \left[\mathrm{ L}_n^\mu(\tilde{L})\right]^2=\frac{(1+\mu)_n \, (-m)_n\,\Gamma(\mu+m+1)}{\Gamma(n+1)^2}\times\quad\quad\quad\quad\nonumber\\
\quad\times\,{}_3F_2(-n,\mu+m+1,m+1;1+\mu,1+m-n;1),\quad \mu> -m-1,\label{calogerosuperformel}
\end{eqnarray}
where $_3F_2(a_1,a_2,a_3;b_1,b_2;z)$ are the generalized hypergeometric functions, defined by
\begin{equation}
_pF_q(a_1,...,a_p;b_1,...,b_q;z)=\sum_{k=0}^\infty\frac{z^k}{k!}\frac{\prod_{i=1}^p (a_i)_k}{\prod_{j=1}^p (b_j)_k},
\end{equation}
and $(a)_n$ are the usual Pochhammer polynomials,
\begin{equation}
(a)_n=\frac{\Gamma(a+n)}{\Gamma(a)}.
\end{equation}
Hence, inserting \eqref{calogerosuperformel} into \eqref{calogeromoments2} gives the final expression for the moments,
\begin{eqnarray}
\expec{L^m}_n&=&\left(\frac{c_1}{4\Lambda}\right)^{\frac{m}{2}} \frac{\Gamma(n-m)\Gamma(\mu+m+1)}{\Gamma(n+1)\Gamma(\mu+1)\Gamma(-m)}\times\nonumber\\
&\times& {}_3F_2(-n,\mu+m+1,m+1;1+\mu,1+m-n;1).
\end{eqnarray}
Note that the poles of $\Gamma(-m)$ cancel with those of the hypergeometric function leading to a finite expression. Further, it is easy to see that all moments scale as $\expec{L^m}_n\!\sim\! \Lambda^{-\frac{m}{2}}$.

\setcounter{equation}{0}
\renewcommand{\theequation}{E.\arabic{equation}}
\section{Discarding unphysical double scaling limits}\label{app:scaling}

In this appendix we discuss certain double scaling limits of the CDT model with topology change, which we discarded as unphysical in Section \ref{sec:topocont}.
In terms of the parameters $\alpha$ and $\beta$ in the scaling relation \eqref{d}, these were the scalings with $\beta\equ1$, instead of the scaling with $\beta\equ\frac{3}{2}$ which led to a physical sensible continuum theory, as discussed in Section \ref{sec:topocont}. 
Hence, for $\beta\equ1$ the scaling relation \eqref{d} reads
\begin{equation}\label{eq:appdis:scaling}
h=\frac{1}{4}\,h_{ren}\,a\,\sqrt{\Lambda}^\alpha (X+Y)^{1-\alpha},
\end{equation}
where the normalization factor on the right-hand side has been chosen 
for later convenience.
To obtain the effective quantum Hamiltonian we follow the same procedure as the one used in Section \ref{sec:topocont}. Upon inserting \eqref{eq:topcont:scalinga}-\eqref{eq:topcont:scaling3} and \eqref{eq:appdis:scaling} into the time evolution of the discrete wave function \eqref{eq:topcont:timeevol}, and expanding both sides to first order in $a$ one obtains
\begin{equation}\label{eq:appdis:expan}
(1-a \hat{H}+\mathcal{O}(a^2))\psi(X) =  \int^{i
\infty}_{-i\infty} \frac{dZ}{2\pi i} \left\{A(X,Z)+B(X,Z) a +
\mathcal{O}(a^2)\right\} \psi(Z),
\end{equation}
where the leading-order contribution is given by
\begin{equation}
A(X,Z)=\frac{2}{(Z-X)\left(1+C(X,Z)\right)}
\end{equation}
with
\begin{equation}
C(X,Z)=\sqrt{1-h_{ren}^2(X-Z)^{-2\alpha}\Lambda^\alpha}.
\end{equation}
For the Laplace transform of $A(X,Z)$ to yield a delta-function, the scaling 
should be chosen such that $\alpha\!\leqslant\! 0$. 
Consider now the terms on the right hand side of \eqref{eq:appdis:expan}, which are of first order in $a$,
\begin{eqnarray}
B(X,Z)&=&\frac{  h_{ren}^2(X+Z-4 Z \gamma ) 
\Lambda ^{\alpha }}{(X-Z)^{1+2\alpha} C(X,Z) \left(1+C(X,Z)\right)^2}\nonumber\\
&-&2\,\frac{X Z-2\Lambda + \gamma  (X-Z)^2 }
{(X-Z)^2 C(X,Z) \left(1+C(X,Z)\right)},\label{app_B}
\end{eqnarray}
one finds that for $\alpha\!\leqslant\! -1$ the continuum limit is independent 
of any new coupling associated to the contribution of the wormholes and therefore
leads to the usual Lorentzian model. This becomes clear when one expands 
the last term of \eqref{app_B} in $(X-Z)$, which yields
\begin{equation}
\frac{X Z-2\Lambda }{(X-Z)^2 C \left(1+C\right)}
=\frac{1}{2}\frac{X Z-2\Lambda }{(X-Z)^2}\left(1+\frac{3}{4}\,h_{ren}^2 \Lambda^\alpha
(X-Z)^{-2\alpha}+\mathcal{O}((X-Z)^{-4\alpha})\right).
\end{equation}
For $\alpha\!\leqslant\! -1$ the term depending on $h_{ren}$ does not have a pole 
and therefore does not contribute to the Hamiltonian.
Since we are only considering non-fractional poles, this leaves as possible
$\alpha$-values only $\alpha=0$ and $\alpha=-\frac{1}{2}$.

\subsection{The case $\beta =1$, $\alpha=0$}

For $\alpha=0$ the Hamiltonian retains a $\gamma$-dependence contained
in the first line of \eqref{app_B}. 
Since there is no immediate physical
interpretation of $\gamma$ in our model, it seems natural to choose
$\gamma=0$, although strictly speaking this does not resolve the
problem of explaining the $\gamma$-dependence of the continuum limit.
Setting this question aside, one may simply look at the resulting model
as an interesting integrable model in its own right. \\

In order to obtain a delta-function to leading order, one still needs
to normalize the transfer matrix by a constant factor $2/(1+s)$,
with $s:=\sqrt{1-h_{ren}^2}$.
After setting $\gamma=0$ and performing an inverse Laplace transformation, the
Hamiltonian reads
\begin{equation}
\label{haml}
\hat{H}(L,\dL)=\frac{1}{s}\left(-L\ddL -s\dL +2\Lambda L \right).
\end{equation}
It is self-adjoint on the Hilbert space $\mathcal{H}=\mathcal{L}^2(\mathbb{R}_+,L^{s-1}dL)$. 
Further setting $L=\frac{\varphi^2}{2\,s}$ one encounters the one-dimensional
Calogero Hamiltonian (cf. Appendix \ref{App:Calogero})
\begin{equation}
\label{calo}
\hat{H}(\varphi,\frac{\partial}{\partial\varphi})=-\frac{1}{2}\frac{\partial^2}{\partial\varphi^2}+
\frac{1}{2}\omega^2\varphi^2-\frac{1}{8}\frac{A}{\varphi^2},
\end{equation}
with $\omega\equ\frac{\sqrt{2 \Lambda}}{s}$ and $A\equ 1-4(1-s)^2$, which implies 
that the model covers the parameter range 
$-3\klgl A\klgl 1$. The maximal range for which the Calogero Hamiltonian is 
self-adjoint is $-\infty \kl A\klgl 1$. It is interesting to see that the usual Lorentzian model without topology changes corresponds 
to $A\equ 1$ in the case of open boundary conditions, and to the value $A\equ -3$ for circular boundary conditions (cf. Appendix \ref{App:Calogero}). This selects the parameter range $-3\klgl A\klgl 1$ as a natural choice.  The Hamiltonian \eqref{calo} has already appeared in a causal dynamically
triangulated model where the two-dimensional geometries were decorated with a
certain type of ``outgrowth" or small ``baby universes" \cite{DiFrancesco:2000nn}.
This model covered the parameter range $0\leqslant A\leqslant 1$.\\

The spectrum of the Hamiltonian \eqref{haml} is given by
\begin{equation}
E_n=\frac{\sqrt{2\Lambda}}{s}(2n+s),\quad n=0,1,2,...\, .
\end{equation}
The corresponding eigenfunctions read
\begin{equation}
\psi_n(L)=\mathcal{A}_n e^{- \sqrt{2\Lambda}L}\,{}_1F_1(-n,s,2\sqrt{2\Lambda} L),
\quad d\mu(L)=L^{s-1}dL,
\end{equation}
where ${}_1F_1(-n,a,b)$ is the confluent hypergeometric function, as defined in \eqref{eq:cont:hypergeom}. 
The eigenfunctions $\{\psi_n(L),n\equ 0,1,2,...\}$ form an orthonormal basis with the normalization factors
\begin{equation}
\mathcal{A}_n=(8\Lambda)^{\frac{s}{4}}\sqrt{\frac{\Gamma(n+s)}{\Gamma(n+1).
\Gamma(s)^2}}
\end{equation}
Let us for completeness also state the finite time partition function or loop-loop correlator
\begin{eqnarray}
	G_{\Lambda,s}(L_1,L_2,T)=\sqrt{2\Lambda}\,(L_1L_2)^{-\frac{s-1}{2}}
	\frac{e^{-\sqrt{2\Lambda}(L_1+L_2)\coth(\frac{\sqrt{2\Lambda}}{s}T)}}{\sinh(\frac{\sqrt{2\Lambda}}{s}T)}
	I_{s-1}\left(\frac{\sqrt{8\Lambda L_1 L_2}}{\sinh(\frac{\sqrt{2\Lambda}}{s}T)}\right).
\end{eqnarray}
with $s\in[0,1]$. One sees explicitly that the case $s\equ 1$ or, equivalently,
$A\equ 1$ corresponds to the pure two-dimensional CDT model with open boundary conditions.

\subsection{The case $\beta =1$, $\alpha=-\frac{1}{2}$}

For $\alpha\equ-\frac{1}{2}$ the result does not depend on $\gamma$ and therefore
on the detailed manner in which we approach the critical point. However, 
the Hamiltonian
\begin{equation}
\hat{H}(L,\dL)=-L\ddL-\dL+2\,\Lambda\,L-\frac{3}{4}\,h_{ren}^2\Lambda^{-1/2}\,\ddL
\end{equation}
cannot be made self-adjoint with respect to any measure $d\mu(L)$ because the 
boundary part of the partial integration always gives a nonvanishing contribution. 
We therefore discard this possibility.

\addcontentsline{toc}{section}{References}

\providecommand{\href}[2]{#2}\begingroup\raggedright\endgroup

\end{document}